\documentclass[journal]{IEEEtran}
\usepackage{amssymb,amsmath,epsfig,psfrag}

\usepackage[english]{babel}
\usepackage{color}
\usepackage[normalem]{ulem}
\usepackage{varioref}

\usepackage{graphicx}
\usepackage{epstopdf}
\usepackage{amsthm}

\newtheorem{theorem}{Theorem}

\newtheorem{Corollary}{Corollary}

\newtheorem{remark}{Remark}

\begin{document}

\title{Effective Capacity of Retransmission Schemes\\
--- A Recurrence Relation Approach}

\author{\IEEEauthorblockN{Peter Larsson \emph{Student
Member, IEEE}, James Gross \emph{Senior
Member, IEEE}, \\ Hussein Al-Zubaidy \emph{Senior
Member, IEEE}, Lars K. Rasmussen \emph{Senior
Member, IEEE}, \\ Mikael Skoglund, \emph{Senior
Member, IEEE.}}%
\thanks{The authors are with the ACCESS Linnaeus Center and the School of Electrical Engineering at  KTH Royal Institute of  Technology, SE-100 44 Stockholm, Sweden.}}
\maketitle


\begin{abstract}
We consider the effective capacity performance measure of persistent- and truncated-retransmission schemes that can involve any combination of multiple transmissions per packet, multiple communication modes, or multiple packet communication. We present a structured unified analytical approach, based on a random walk model and recurrence relation formulation, and give exact effective capacity expressions for persistent hybrid automatic repeat request (HARQ) and for truncated-retransmission schemes. For the latter, effective capacity expressions are given for systems with finite (infinite) time horizon on an algebraic (spectral radius-based) form of a special block companion matrix. In contrast to prior HARQ models, assuming infinite time horizon, the proposed method does not involve a non-trivial per case modeling step. We give effective capacity expressions for several important cases that have not been addressed before, e.g. persistent-HARQ, truncated-HARQ, network-coded ARQ (NC-ARQ), two-mode-ARQ, and multilayer-ARQ. We propose an alternative QoS-parameter (instead of the commonly used moment generating function parameter) that represents explicitly the target delay and the delay violation probability. This also enables closed-form expressions for many of the studied systems. Moreover, we use the recently proposed matrix-exponential distributed (MED) modeling of wireless fading channels to provide the basis for numerous new effective capacity results for HARQ. 
\end{abstract}


\section{Introduction}
\label{sec:Intro}
\IEEEPARstart{M}{odern} wireless communication systems, such as cellular and WLAN systems, are data packet oriented services operating over unreliable channels that typically target reliable high data-rate communication. In order to achieve this goal, most wireless systems employ some form of retransmission scheme. Commonly,  retransmission schemes are classified (according to their functionality) either as automatic repeat request (ARQ), or as hybrid-ARQ (HARQ) \cite{bibLin}, \cite{bibLin2}. HARQ, with soft combining of noisy redundancy-blocks in error, is often employed in wireless communication systems. This kind of HARQ scheme, to be discussed in Section \ref{sec:RetrSchemes}, is further classified into repetition-redundancy-HARQ (RR)\footnote{Often called Chase combining in past literature, but we opt for the naming convention RR-HARQ as discussed in \cite{bibLarsson1}.}  \cite{bibSindhu}-\cite{bibChase}, and incremental-redundancy-HARQ (IR) \cite{bibMandelbaum}. Another line of work considers truncated-HARQ, which imposes an upper limit on the number of retransmission attempts, a \textit{transmission limit}, of HARQ. This was introduced in \cite{bibFujiwara}, and investigated further in \cite{bibYang}. A shared feature of these schemes is the (potential) use of multiple transmissions to send a packet. Other more recently proposed retransmission principles are network coded-(H)ARQ (NC-(H)ARQ) \cite{bibLarsson3,bibLarsson4} (using network coding, e.g. bit-wise XOR-ing amid packets, with (H)ARQ) and multilayer-(H)ARQ \cite{bibShamai}-\cite{bibLarsson6} (using super position coding with (H)ARQ). They share another throughput-enhancing feature, namely that multiple packets can be communicated concurrently. NC-(H)ARQ has yet another interesting feature, it alters between different communication modes (when sending regular- or NC-packets). Moreover, schemes shifting between different channel states can also be seen as changes in communication modes \cite{bibBoujemaa}. This selection of works studies one particular retransmission scheme at a time. We believe that it would be useful with a structured approach that allows modeling and analysis of a general class of retransmission schemes which is characterized by multiple transmissions, multiple packets, and multiple communications modes. 

\begin{figure}[t]
 \centering
 \vspace{+.1 cm}
 \includegraphics[width=10cm]{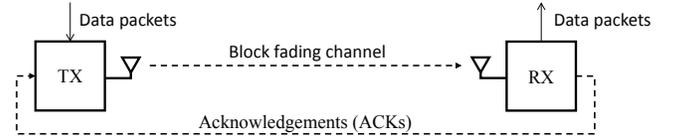}
 \vspace{-5.6cm}
 \caption{Communication system with retransmissions.}
 \label{fig:System}
 \vspace{-0.3cm}
\end{figure}

Typically, the throughput performance of retransmission schemes are evaluated and analyzed. In \cite{bibWang}, \cite{bibZorzi}, the throughput of (H)ARQ was defined and studied based on renewal theory. With such definition, an information theoretical approach for analyzing HARQ was established in \cite{bibCaire}. Numerous works on (H)ARQ throughput analysis, e.g \cite{bibBettesh}-\cite{bibLarsson2}, have adopted this information theoretical approach.
However, for some data services, such as video-streaming, quality-of-service (QoS) guarantees in terms of bounded delays are often more desirable. Unfortunately, the throughput metric is not well suited for this purpose. An alternative performance metric for (queuing) systems with varying service rates, e.g., data transmission over fading channels, and delay targets, is the notion of effective capacity. This metric, introduced in \cite{bibWu}, was inspired by the large deviation principle and the concept of effective capacity, \cite{bibChang, bibKelly}. The objective of the effective capacity is to quantify the maximum sustainable throughput under stochastic QoS guarantees with varying server rate, but it also allows probabilistic delay targets to be considered. 

Using the information theoretical mentioned above, the effective capacity metric has been used to study many wireless systems, \cite{bibWu}-\cite{bibAkin2}. The use of adaptive modulation and coding (AMC) with signal-to-noise-ratio (SNR) dependent rate adaptation was analyzed in \cite{bibWu, bibChoi, bibTang2}, whereas ARQ was considered in \cite{bibChoi, bibLi, bibHarsini, bibAkin2}, and joint AMC and ARQ was studied in \cite{bibTang1, bibMusavian, bibFemenias}. ARQ with non-capacity achieving Reed-Solomon codes was studied in \cite{bibFidler}. The subject of cooperative-ARQ for relays was analyzed in \cite{bibHu, bibHarsini}. Of more relevance, for this work, are the HARQ-like systems addressed in \cite{bibLi, bibGunaseelan, bibAkin2}. On the modeling side, finite state Markov chains (FSMC) were used for modeling the effects of channel variations in \cite{bibTang1}, \cite{ bibGunaseelan}-\cite{ bibAkin2}.
The work \cite{bibTang2} considers such FSMC-modeling for AMC with channel fading but do not consider any retransmission scheme. Also, FSMC-modeling for joint AMC and ARQ with channel fading is used in \cite{bibTang1, bibFemenias} which assumes that a large number of ARQ cycles in each AMC state, each modeled as a packet erasure channel. Hence, for one AMC state, the effective capacity equals that of throughput. This reveals that the operation differs from ARQ modeling with a reward on successful, and none on failed, transmission. 
The combined effects of channel fading and IR-HARQ (with two transmission attempts) are approximated, using a finite state Markov process, in \cite[Sec. II.B, Fig. 2]{bibGunaseelan}. The underlying assumption (and problem) for this approximative model is the same as in \cite{bibTang1, bibFemenias}. Also note  in \cite[Fig. 2]{bibGunaseelan}, unlike a realistic model of truncated-IR, a direct transition from a failed second transmission attempt to a successful first transmission is prohibited. 
Similarly to this work, HARQ with truncated transmissions was considered in \cite{bibAkin2}. Yet, the mathematical modeling in \cite[(9)]{bibAkin2} implies that packets are always correctly received on the last transmission attempt. In this work, in contrast, a packet is discarded, as in a real system, if it fails on the last transmit attempt. Onwards, we refer to the scheme in \cite{bibAkin2} as "guaranteed-success"-truncated-HARQ.
We note that \cite{bibWu}-\cite{bibAkin2} assumed Rayleigh fading channels, whereas \cite{bibTang1, bibMusavian} also considered Nakagami-$m$ fading.

We conclude that (i) no works have analyzed and given the exact effective capacity of persistent/truncated-HARQ and NC-ARQ, (ii) no structured effective capacity methodology for handling the general class of multi-transmission, -packets, and communication modes has been proposed, (iii) closed-form effective capacity expression are rare, and analytical effective capacity optimization results are not known, and (iv) existing studies are (almost) exclusively limited to Rayleigh fading.

\subsection{Contributions}
\label{sec:Contributions}
The contributions can be divided into three levels, (i) a methodology, (ii) results for important (H)ARQ schemes, and (iii) methods for more useful, general, closed-form results.

On the first level is a structured and unified method for effective capacity analysis of the general class of retransmission schemes involving multiple-transmissions, -packets, and -communication modes. The operation of such schemes is modeled as a three-dimensional random walk (3D-RW). This class is fully defined by a set of transition probabilities indexed in terms of transmissions, packets, and modes. The effective capacity is determined by solving a certain recurrence relation, yielding a matrix- and a characteristic equation-based solution.
Considering the matrix solution case now. In related works, e.g. \cite{bibTang2}-\cite{bibAkin2}, the systems are modeled as Markov modulated processes, each defined by a transition probability matrix $\mathbf{P}$ and a diagonal reward matrix $\boldsymbol{\Phi}$, for which the effective capacity is computed from the spectral radius of $\mathbf{P}\boldsymbol{\Phi}$. 
The extension of this approach to the analysis of the general retransmission class defined and considered in this work may not be feasible since the approach involves a non-trivial per case modeling step, i.e., the contribution of any such work will be the modeling of the system by a Markov modulated process which can only be performed per individual retransmission scheme. Due to the difficulty involved in the modeling step, which incidentally is highly dependent on the skill level of the researcher/engineer  crafting that model, this approach remains case-specific, limited to 'easy to model' schemes and error-prone (due to dependency on the modeler skill level) and the resulting Markov chain model is not unique since it depends on the modeler interpretation of the system. This is evident from the fact that even when spectral radius based solutions of the $\mathbf{P}\boldsymbol{\Phi}$ type models existed for several decades now, very few works in the literature used this methodology to analyze retransmission systems. One such example is \cite{bibAkin2} which uses, as discussed in the related work section, the unrealistic simplifying assumption that the last permitted retransmission attempt in a truncated-ARQ scheme is always successful in order to make the model tractable. In contrast, our proposed approach avoids such Markov chain modeling step by providing a systematic way to compute the companion matrix entries (see equation (\ref{eq:amss})) for any number of transmission states, packets per transmission, and transmission modes.
This method also leads to resulting model matrices of lower dimension (determined only by the number of transmission attempts and communication modes, but independent of the number of reward rates), less complex forms (without, e.g., transition probabilities repeated in multiple matrix entries and potential ratios of transition probabilities), and lower computational complexity compared to the Markov modulated approach. We also believe that the methodology is more intuitive and thus easier to apply. This simplicity cater for simpler analysis, and insights to be gained, for any existing and hypothesized retransmission scheme.

On the second level, other significant and novel contributions of this work are the effective capacity expressions for truncated-HARQ, persistent-HARQ (which relies on the characteristic equation solution), and NC-ARQ. Effective capacity expressions of any sort have not been reported for these systems prior to this work.

On the third level, other contributions are effective capacity analysis and expressions, for systems, with $k$-timeslots, a joint parametrization of target delay and delay violation probability, and matrix exponential distributed (MED) fading channels.

A detailed list of the main contributions are:
\begin{itemize}
\item A structured and unified effective capacity analysis, with exact expression(s), of multi-transmission, -mode, and -packet truncated-retransmission schemes: Theorem \ref{theorem:CeffRetrScheme} (Corollary \ref{Corollary:CeffRetrSchemekInf}) for finite (infinite) number of timeslots.

\item Effective capacity expression of persistent-HARQ: Theorem \ref{theorem:ECHARQEffCh}, and Corollaries \ref{Corollary:CpsiRR}-\ref{Corollary:CpsiRRRayleigh}.

\item A closed-form effective capacity expression of classical truncated-HARQ (i.e. with a packet discard on the last transmission if the final decoding effort fails): Corollary \ref{Corollary:CeffHARQ}. A closed-form expression for $M=2$: Corollary \ref{Corollary:Ceff2}.

\item An effective capacity expression of network-coded ARQ: Corollary \ref{Corollary:NCARQ}.

\item Closed-form effective capacity expressions given in terms of a proposed (more practical) QoS-parameter $\psi$, dependent on the target delay and the delay violation probability: Corollaries \ref{Corollary:CpsiHARQ1}-\ref{Corollary:CpsiMEDRRopt}, \ref{Corollary:CpsiRRRayleigh}, and \ref{Corollary:CpsiARQ1}.
\item Closed-form effective capacity expressions of (H)ARQ schemes for fading channels formulated with the matrix exponential distribution: Corollaries 
\ref{Corollary:CpsiMEDRR}, \ref{Corollary:CpsiMEDRRopt}, and in part \ref{Corollary:CpsiRRRayleigh},  \ref{Corollary:CpsiARQ1}.
\end{itemize}
Additional detailed contributions of the paper are:
\begin{itemize}  
\item A closed-form expression for $M=2$: Corollary \ref{Corollary:Ceff2}.
\item A closed-form effective capacity expression, and its optimization, of persistent-HARQ expressed in the QoS-parameter $\psi$ (including delay target and delay violation probability), and the versatile MED effective channel: Corollaries \ref{Corollary:CpsiMEDRR} and \ref{Corollary:CpsiMEDRRopt}.
\item A closed-form effective capacity expression expressed in the QoS-exponent $\theta$ for RR in block Rayleigh fading: Corollary (\ref{Corollary:SISORR}).
\item An effective capacity expression of two-mode-ARQ in general, and of ARQ in a Gilbert-Elliot block fading channel in particular: Corollary \ref{Corollary:TwoModeARQ} and Section \ref{sec:2modeARQ}.
\item A new effective capacity approximation of HARQ (expressed in first and second moments): Corollary \ref{Corollary:CeffAppr}.
\item A recurrence relation for the $\alpha$-moment, $\mathbb{E}\{N_k^\alpha\}$, and a $k$-timeslot throughput expression: Corollary \ref{Corollary:THARQkSlot}.
\end{itemize}

\subsection{Organization}
\label{sec:Organization}
\begin{figure}[t]
\centering
\includegraphics[width=9cm]{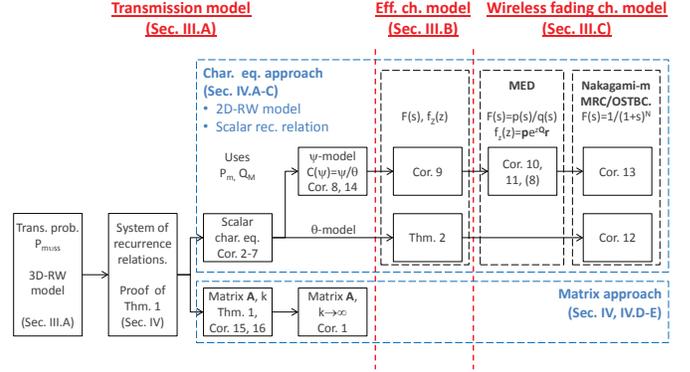}
\vspace{-2.3cm}
\caption{Roadmap of paper}
\label{fig:T_1T_4}
\vspace{-0.4cm}
\end{figure}

The paper is organized as follows. In Section \ref{sec:Preliminaries}, we introduce the notation, the retransmission schemes, and the effective capacity measure. The (three-level hierarchical) system model with the random walk model is described in Section \ref{sec:SystemModel}. In Section \ref{sec:ECofRetrSchemes}, we first derive general effective capacity expression(s), and then specialize to truncated/persistent-HARQ (with the three levels of the system model), NC-ARQ, and two-mode ARQ. Numerical and simulation results are presented along with the studied cases. In Section \ref{sec:Conclusions}, we summarize and conclude. For the readers convenience, we also illustrate a roadmap for the analysis progression of this work in Fig. \ref{fig:T_1T_4}.

\section{Preliminaries}
\label{sec:Preliminaries}
We depict the communication system in Fig. \ref{fig:System} with one transmitter, one receiver, and feedback. Data packets that arrive to the transmitter are sent over the (unreliable channel) and then forwarded by the receiver. As the packet arrives, they are queued, if needed, until a communication opportunity appears. For the communication part, data packets are channel-encoded into codewords (or incremental redundancy block) of initial rate $R$ [b/Hz/s]. One may think of $N_\textrm{b}$ [bits] uncoded bits being sent over a bandwidth $B_\textrm{Hz}$ [Hz], and for a duration $D_\textrm{s}$ [s], which gives rate $R\triangleq {N_\textrm{b}}/{B_\textrm{Hz}D_\textrm{s}}$. Note that the rate $R$ remains fixed for each retransmission. At the receiver side, channel decoding takes place where the receiver may, depending on retransmission scheme considered, exploit stored information acquired from past communication attempts. After channel decoding (and potentially other processing) at the receiver side, the receiver acknowledges if a data packet is deemed error-free, or not. Below, we introduce the notation, give more details on the retransmission schemes, and discuss the effective capacity performance metric.

\subsection{Notation and Functions}
\label{sec:Notation}
We let $x$, $\mathbf{x}$, and $\mathbf{X}$ represent a scalar, a vector, and a matrix, respectively. The transpose of a vector, or a matrix, is indicated by $(\cdot)^T$. $\mathbb{P}\{X=x\}$ denotes the probability that a random variable (r.v.) $X$ assumes the value $x$, whereas $\mathbb{E}\{X\}$ is the expectation value of $X$. A probability density function (pdf) is written as $f_X(x)$. We further let $f^{\circledast(k)}(x)$ denote the $k$-fold convolution. The Laplace transform \cite[17.11]{bibGradshteyn}, with argument $s$, is written as $\mathcal{L}_s \{\cdot\}$, whereas the inverse Laplace transform, with argument $x$, uses the notation $\mathcal{L}_x^{-1} \{\cdot\}$. For functions, $W_{0}(x)$ is the principal branch ($W_0(x)>-1$) of Lambert's $W$-function, defined through $x=W(x)\mathrm{e}^{W(x)}$ \cite{bibCorless}. The regularized lower incomplete gamma function is  $\gamma_\textrm{r}(k,x)\triangleq\frac{1}{(k-1)!}\int_0^x t^{k-1}\mathrm{e}^{-t} \, \mathrm{d}t$.

\subsection{Retransmissions Schemes}
\label{sec:RetrSchemes}
One kind of classification is based on how the information bearing signal is composed, sent and then processed at the receiver side. In this respect, the fundamental types are HARQ and ARQ. In HARQ, the receiver exploits received (and stored) information from past transmission attempts to increase the probability of successful decoding of a data packet. In ARQ, the receiver exploits no such information, and each transmit attempt is seen as a new independent communication effort. Retransmission schemes are traditionally evaluated in terms of their throughput. The throughput is defined as the ratio between the mean amount of delivered information of a packet, and the mean number of transmissions of a packet. Using this definition, and allowing for at most $M$ transmissions per data packet, the throughput of truncated-(H)ARQ is known, \cite{bibCaire}, to be 
\begin{align}
T_\textrm{trunc.}^\textrm{HARQ}\triangleq \frac{R(1-Q_M)}{\sum_{m=1}^M m P_m+MQ_M}=\frac{R(1-Q_M)}{\sum_{m=0}^{M-1} Q_m}.
\label{eq:THARQtrunc}
\end{align}
In (\ref{eq:THARQtrunc}), $P_m$ is the probability of an error-free packet decoded at the $m$th transmission, $Q_m$ is the probability of failed decoding of a data packet up to and including the $m$th transmission. Hence, we can write $P_m=Q_{m-1}-Q_m$, $Q_0\triangleq1$. For ARQ, with a memoryless iid channel, we see that $P_m^\textrm{ARQ}=Q_{m-1}-Q_m=(1-Q_1)Q_1^{m-1}=P_1Q_1^{m-1}$. Inserting, $P_m^\textrm{ARQ}$ in (\ref{eq:THARQtrunc}), we find that $T_\textrm{trunc.}^\textrm{ARQ}=RP_1$, irrespective of transmission limit $M$. In HARQ, as the receiver exploits information from past transmission attempts, $P_m^\textrm{HARQ}\geq P_m^\textrm{ARQ}, m\geq2$. Now, letting $M\rightarrow \infty$, we get the throughput for persistent-(H)ARQ
\begin{align}
T_\textrm{persistent}^\textrm{HARQ}\triangleq\frac{R}{\sum_{m=1}^\infty mP_m}=\frac{R}{\sum_{m=0}^\infty Q_m},
\label{eq:THARQloss}
\end{align}
which acts as an upper bound to the throughput of truncated-HARQ. In wireless communication systems, RR- and IR-HARQ are frequently used. In RR/IR-truncated-HARQ, redundancy blocks are transmitted up to the point that a packet is correctly decoded, or $M$ attempts have been made. For RR, redundancy blocks are merely repetitions of the channel coded data packet, and for IR, the redundancy blocks are channel code segments derived from a low-rate code word. The processing at the receiver side for RR involves maximum ratio combining (or interference rejection combining in the presence of interference) and subsequent channel decoding, whereas for IR the receiver jointly channel-decodes all redundancy blocks received for a data packet still in error.  As noted, ARQ has the throughput $T^\textrm{ARQ}_\textrm{trunc.}=RP_1$. However, when serving multiple receivers with individual data, and soft information from previous transmission is not stored, basic ARQ does not give the highest throughput. For this scenario, NC-ARQ yields higher throughput. The core idea is to send network-coded packets to users that have overheard each other's past transmissions. For a two-user system, the throughput in a symmetric packet erasure channel has been found to be $T^\textrm{2NCARQ}=R2P_1(2-P_1)/(3-P_1)\geq R P_1$, \cite{bibLarsson3}. To model channels with memory, a block Gilbert-Elliot channel, altering between a good and a bad channel state is a simple but useful option. For this case, the throughput for ARQ is $T^\textrm{GE-ARQ}=P_\textrm{GG}T_\textrm{GG}^\textrm{ARQ}+P_\textrm{BB}T_\textrm{BB}^\textrm{ARQ}$,  for $P_\textrm{GG}=1-P_\textrm{BB}$ \cite{bibBoujemaa}, where GG and BB represents being in the good and the bad channel state, respectively. NC-(H)ARQ, as well as ARQ in a Gilbert-Elliot channel, are examples of where the retransmission scheme operate in different communication modes over time. Most retransmission schemes deliver only a fixed amount of information of rate $R$ at each communication instance. In contrast, NC-ARQ can communicate multiple packets concurrently. Multilayer-ARQ, can also send multiple packets at the same time. This is accomplished by transmitting a superposition of codewords, with rates  $\{r_1,r_2,\ldots,r_L\}$ and fractional power levels  $\{x_1,x_2,\ldots,x_L\}$, and exploiting the possibility that multiple codewords can concurrently be correctly decoded depending on the channel fading state.

\subsection{Effective Capacity}
\label{sec:EC}
\begin{figure}[t]
 \centering
 \vspace{+.1 cm}
 \includegraphics[width=10cm]{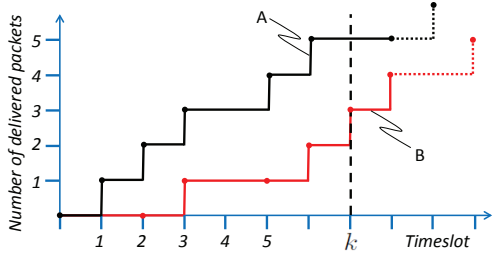}
 \vspace{-4.1cm}
 \caption{Examples, A and B, of transmission event sequences, for $M=2$ truncated-HARQ.}
 \label{fig:Servicecurve}
 \vspace{-0.3cm}
\end{figure}

Using the effective bandwidth framework in \cite{bibChang}, the concept of effective capacity was proposed in \cite{bibWu}. The effective capacity corresponds to the maximum sustainable source rate, and is (typically) defined as the limit
\begin{align}
C_\textrm{eff}&\triangleq -\lim_{k\rightarrow \infty} \frac{1}{\theta k}\ln \left( \mathbb{E}\left\{ \mathrm{e}^{-\theta \zeta_k}\right \} \right),
\label{eq:Ceff}
\end{align}
where $\zeta_k$ is the accumulated service process at time $k$, and $\theta$ is the so called QoS-exponent. In general, $\zeta_k$ can assume continuous or discrete values, depending on the serving channel. However, a more general definition of the effective bandwidth, for finite time $k$, was considered in \cite{bibKelly}. This suggests an alternative, more general, effective capacity definition
\begin{align}
C_{\textrm{eff},k}&\triangleq - \frac{1}{\theta k}\ln \left( \mathbb{E}\left\{ \mathrm{e}^{-\theta \zeta_k}\right \} \right),
\label{eq:Ceffk}
\end{align}
which reflects a system with a $k$-timeslot long window for communication\footnote{Later, we show that this can also be motivated with respect to throughput measured over a $k$-timeslot long window.}. The notion of effective capacity also allows for determining the maximum fixed source rate under a statistical QoS-constraint,  $\mathbb{P}\{D>D_\textrm{max}\}\leq \epsilon$, where $D$ is the steady state delay of packets in the source queue, $D_\textrm{max}$ is the delay target, and $\epsilon$ is the limit of the delay violation probability. This connects back to the original motivation in Section \ref{sec:Intro}, on QoS-enabled performance metrics. In \cite{bibHarsini}, it was shown that, when $D_\textrm{max}\rightarrow \infty$, the following holds
\begin{align}
\mathbb{P}\{D>D_\textrm{max}\}\simeq \eta \mathrm{e}^{-\theta C_\textrm{eff}(\theta)D_\textrm{max}},
\label{eq:ProbDmax}
\end{align}
where $\eta$ is the probability that the queue is non-empty. Combining the QoS-constraint, (\ref{eq:ProbDmax}), and rearranging, we find the QoS-exponent $\theta^*$ of interest by solving
\begin{align}
\theta^* C_\textrm{eff}(\theta^*)= \frac{\log{(\eta/\epsilon)}}{D_\textrm{max}}\triangleq \psi.
\label{eq:thetaC}
\end{align}
Note here that we also define a QoS-parameter $\psi$, that jointly reflects the delay target, the delay violation probability, and the probability of a non-empty queue, in (\ref{eq:thetaC}). Thus, the maximum source rate under a statistical QoS-constraint,  $\mathbb{P}\{D>D_\textrm{max}\}\leq \epsilon$, is  $ C_\textrm{eff}(\theta^*)$.
For HARQ, either a packet is in error, or a packet of rate $R$  is communicated error-free. Therefore, (\ref{eq:Ceff}) becomes
\begin{align}
C_\textrm{eff}^\textrm{HARQ}&\triangleq -\lim_{k\rightarrow \infty} \frac{1}{\theta k}\ln \left( \mathbb{E}\left\{ \mathrm{e}^{-\theta R N_k}\right \} \right)\notag\\
&=-\lim_{k\rightarrow \infty} \frac{1}{\theta k}\ln \left( \sum_{\forall n} \mathrm{e}^{-\theta R n} \mathbb{P}\{N_k=n\} \right),
\label{eq:CeffNk}
\end{align}
where $N_k$ is the accumulated service process in number of correctly decoded packets at time $k$. We illustrate two example service processes (or transmission event sequences) in Fig. \ref{fig:Servicecurve} for  truncated-HARQ. We note that at time $k$, the service process can terminate, or pass by, at time $k$. It is, in part, this aspect that makes the analysis challenging. For the case where the transmissions are independent, as is the case for ARQ, the effective capacity simplifies to
\begin{align}
C_\textrm{eff}^\textrm{ARQ}&\triangleq -\ \frac{1}{\theta }\ln \left( \mathbb{E}\left\{ \mathrm{e}^{-\theta R N_1}\right \} \right)\notag\\
&=- \frac{1}{\theta }\ln \left( Q_1+ \mathrm{e}^{-\theta R} P_1 \right),
\label{CeffARQ}
\end{align}
where  $P_1=\mathbb{P}\{N_1=1\}$ and $Q_1=\mathbb{P}\{N_1=0\}=1-P_1$.

\section{System Model}
\label{sec:SystemModel}
The system model is hierarchical, and contains three levels with increasing specialization, a transmission model, an effective channel model, and a wireless fading channel model.

\subsection{Transmission Model}
\label{sec:SystModel1}
\begin{figure}[t]
 \centering
 \vspace{+.1 cm}
 \includegraphics[width=10cm]{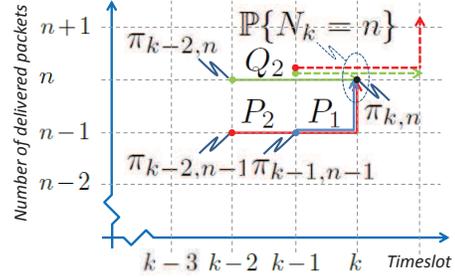}
 \vspace{-3.5cm}
 \caption{Example of transmission model for $M=2$ truncated-HARQ, with $S=1$, $\nu=\{0,1\}$, $P_1=P_{1111}$, $P_2=P_{2111}$ and $Q_2=P_{2011}$.}
 \label{fig:Transitions}
 \vspace{-0.3cm}
\end{figure}

A useful depiction of the proposed retransmission system model is as a (constrained) three-dimensional random-walk (3D-RW) model with varying step sizes on a grid.  Each random 3D-step represents a transmission cycle and is characterized by a \textit{transition probability} $P_{m\nu \tilde s s}$, where $m$ represents the number of transmissions needed until successful decoding, $\nu$ is the number of  correctly decoded packets, $\tilde s$ is the originating communication mode, and $s$ is the final communication mode during a retransmission cycle. This intuitive, and naturally formulated, model, with values for $P_{m\nu \tilde s s}$, fully characterize any retransmission scheme and its performance.
With communication modes, we mean that a retransmission scheme can shift between different (re-)transmission strategies or communication conditions. For example, in NC-ARQ, Section \ref{sec:NCARQ}, the sender alternate between sending regular and network-coded packets, and for 2-mode ARQ, Section \ref{sec:2modeARQ}, the channel statistics alternates. 
More formally, we assume that we start a transmission cycle at a \textit{transmission event state} $\{\tilde{k}, \tilde{n}, \tilde{s}\}$, where $\tilde{k}$ is the current timeslot, $\tilde{n}$ is the current number of correctly decoded packets, and $\tilde{s}$ is the current communication mode. At the end of the transmission cycle, we assume that the transmission event state will be $\{k,n,s\}$. Hence, a transmission cycle lasts $m \triangleq k-\tilde{k}$ transmissions to complete, $\nu \triangleq  n-\tilde{n}$ number of error-free packets are communicated, and the communication mode changes from $\tilde{s}$ to $s$. We limit the model to the ranges $m=\{1,2,\ldots, M \}$, $\nu=\{0,1,\ldots, \nu_\textrm{max}\}$, $s=\{1,2,\ldots, S\}$, and $\tilde{s}=\{1,2,\ldots, S\}$. As before, $M$ is the maximum number of transmission attempts, $\nu_\textrm{max}$ is the maximum number of packets (each of rate $R$) that can be communicated in a transmission cycle, and $S$ is the number of communication modes.
We note that the sum of all transition probabilities exiting a transmission event state $\{\tilde{k}, \tilde{n}, \tilde{s}\}$ adds to one for each $\tilde{s}$. We let $\pi_{k,n,s}$ denote the \textit{recurrence termination probability} of all \textit{transmission event sequences} terminating in $\{k,n,s\}$. We further let $\mathbb{P}\{N_k=n,S_k=s\}$ signify the \textit{state probability} of all transmission event sequences either terminating in, or passing through, $\{k,n,s\}$.
In Fig. \ref{fig:Transitions}, we illustrate the transition probabilities, the recurrence termination probabilities, and the state probabilities, for truncated-HARQ with $M=2$. Fig. \ref{fig:Servicecurve} and \ref{fig:Transitions} also illustrate truncated-HARQ as a 2D-RW.
In addition, we also assume ideal retransmission operation with error-free feedback, negligible protocol overhead, ideal error detection with zero probability of mis-detection, and other standard simplifying assumptions.
We now consider truncated-HARQ, as an example of schemes captured by the transmission model and the notation. For truncated-HARQ, we have $S=1$, $\nu=\{0,1\}$, and the probability of successful decoding on the $m$th transmission attempt relates to the transition probabilities as $P_m\triangleq P_{m111}, m\in\{1,2\ldots,M\}$. Likewise, the probability of failing to decode on the $M$th transmission relates as $Q_{M}\triangleq P_{M011}$.

\subsection{Effective Channel Model}
\label{sec:SystModel2}
The effective channel, a model motivated and explored in \cite{bibLarsson1} for throughput analysis of HARQ, is assumed fully characterized by a pdf $f_Z(z)$, its Laplace transform $F(s)$, and a decoding threshold $\Theta$. A short motivation is given below. Using the notion of information outage, the probability of failing to decode, all up to and including, the $m$th transmission for a capacity achieving codeword is $Q_m\triangleq \mathbb{P} \{ C_m<R \}$, where $C_m$ denotes the cumulative mutual information up to the $m$th transmission. For HARQ, we can typically write $Q_m$ on the form $ Q_m= \mathbb{P} \{ \sum_{{m'}=1}^m z_{m'}<\Theta \}$, where $\sum_{{m'}=1}^m z_{m'}$ is a sum of $m$ iid r.v.s $z_{m'}$ (representing the effective channel), and $\Theta$. We illustrate this with two examples. For RR in block fading channel, with SNR $\Gamma$ and unit variance channel gain $g_{m'}$ for the $m'$th transmission, we have $Q_m^\textrm{RR}= \mathbb{P} \{ \log_2 (1+\Gamma \sum_{{m'}=1}^m g_{m'} )<R \}=\mathbb{P} \{ \sum_{{m'}=1}^m g_{m'}<({2^R-1})/{\Gamma} \}$. This gives $z_{m'}^\textrm{RR}=g_{m'}$, and $\Theta^\textrm{RR}=({2^R-1})/{\Gamma}$. In a similar fashion for IR, we get $Q_m^\textrm{IR}\triangleq \mathbb{P} \{ \sum_{{m'}=1}^m \log_2 (1+\Gamma  g_{m'})<R \}$, which implies $\Theta^\textrm{IR}=R$, and $z_{m'}^\textrm{IR}= \log_2(1+\Gamma  g_{m'})$. For an effective channel, it is assumed that $z_{m'}$ and $\Theta$ reflect the combination of HARQ scheme, channel fading statistics, modulation and coding scheme, rate $R$, and SNR $\Gamma$. The decoding failure probability can also be expressed in $F(s)$ as
\begin{align}
Q_m&=\int_0^\Theta f_Z^{\circledast(m)}(z) \, \mathrm{d}z = \int_0 ^\Theta \mathcal{L}_z^{-1}\{F(s)^m\} \, \mathrm{d}z.
\label{eq:Qm}
\end{align}

\subsection{Wireless Fading Channel Models}
\label{sec:SystModel3}
Here, we use the MED model, and notation, introduced in \cite{bibLarsson2} for throughput analysis of HARQ. The MED has pdf $f_Z(z)=\mathbf{p}\mathrm{e}^{z\mathbf{Q}}\mathbf{r}$, $\mathbf{Q}=\mathbf{S}-\mathbf{r}\mathbf{q}$, where $\mathbf{p}$ and $\mathbf{q}$ are MED-parameter row vectors, $\mathbf{S}$ is a shift matrix (a square all-zero matrix except ones on the superdiagonal) with the dimensions given by the context, and $\mathbf{r}=[0 \ 0 \ldots \ 1]^T$. The MED is equivalent to a sum of exponential-polynomial-trigonometric terms, and has a rational Laplace transform $F(s)=p(s)/q(s)$, where $p(s)$ and $q(s)$ are polynomials, with $\deg(q(s))>\deg(p(s))$ and $q(s)$ is monic. The entries in $\mathbf{p}$ and $\mathbf{q}$ corresponds to the coefficients of $p(s)$ and $q(s)$, respectively. An important idea conveyed in \cite{bibLarsson2} is that the MED can compactly represent a broad class of (effective) wireless fading channels, all with a rational $F(s)$, such as, but not limited to, Nakagami-$m$- and Rayleigh-fading with maximum ratio combining and/or selection diversity. To exemplify, consider orthogonal space-time-block (STC) coding (with STC rate $R_\textrm{stc}$, $N_\textrm{t}$ transmit antennas, and $N_\textrm{r}$ receive antennas) and maximum-ratio-combining, operating in an iid block Nakagami-$m$ fading channel (with parameter $m^\textrm{N}$). 
Based on the model in \cite{bibLarsson1}, we get $Q_m=\gamma_\textrm{r}(m\tilde N, \tilde \Theta)$,$\tilde \Theta=(2^{\tilde R}-1)/\tilde{\Gamma}$, $\tilde R={R/R_\textrm{stc}}$, $\tilde{\Gamma}=\Gamma/R_\textrm{stc}N_\textrm{t}$, $\tilde N=N_\textrm{t}N_\textrm{r}m^\emph{N}$, and the Laplace transform of the effective channel is then a MED channel $F(s)=1/(1+s)^{\tilde N}$. For RR and a Rayleigh fading channel, this simplifies to $\tilde \Theta=(2^R-1)/\Gamma$ with $F(s)=1/(1+s)$.

\section{Effective Capacity of Retransmission Schemes}
\label{sec:ECofRetrSchemes}
In this section, we consider the effective capacity of a general class of truncated-retransmission schemes (with potential packet discards on the last transmission attempt), allowing for multiple-transmissions, multiple-modes, and multiple-packets. Subsequently, we specialize the overall analysis and results to HARQ, illustrating the case of multiple-transmissions, and then (for example) to NC-ARQ, illustrating the case of multiple-modes and -packets. For the initial general analysis, we start by studying the effective capacity of a finite $k$-slotted retransmission system in Theorem \ref{theorem:CeffRetrScheme}, and then proceed to develop the analysis for the limiting case, with infinite $k$.
\begin{theorem}
\label{theorem:CeffRetrScheme}
The effective capacity of a retransmission scheme, with $k$ timeslots, $M$ as transmission limit, $S$ communication modes, $\nu\in\{0,1,\ldots,\nu_\emph{max}\}$ packets per transmission cycle, and the probability of successful decoding on the $m$th attempt $P_{m \nu \tilde{s} s}$, is
\begin{align}
C_{\emph{eff},k}^\emph{Retr.}
&=-\frac{1}{\theta k} \ln{\left(\mathbf{b}^T\mathbf{A}^{k-M} \mathbf{f}_{M}\right)}, \, k\geq M
\label{eq:CeffRetrk}
\end{align}
where $\mathbf{b}=[\mathbf{1}^{1\times S} \ \mathbf{0}^{1\times S(M-1)}]^T$, $\mathbf{f}_M$ is an initial vector\footnote{It is also possible to express (\ref{eq:CeffRetrk}) on the form $C_{\textrm{eff},k}^\textrm{Retr.}=-\frac{1}{\theta k} \ln{\left(\mathbf{b}^T\mathbf{f}_{k}\right)}$, where $ \mathbf{f}_{k}=\mathbf{A} \mathbf{f}_{k-1}+\mathbf{c}_{k-1}, \, k\geq 1$. $\mathbf{f}_{0}$ represents the initiation vector, e.g. $\mathbf{f}_{0}=[1 \ \mathbf{0}]^T$. We find that $c_{k,\tilde{s}} =\sum_{m=k+1}^M\sum_{s=1}^S\sum_{\nu=0}^{\nu_\textrm{max}} P_{m\nu \tilde{s}s}$, where $\mathbf{\tilde{c}}_{k}=
\begin{bmatrix} 
c_{k,1} & c_{k,2} & \ldots & c_{k,S} 
\end{bmatrix}^T$,
and 
$\mathbf{c}_{k}=
\begin{bmatrix} 
\mathbf{\tilde{c}}_{k} & \mathbf{0} & \ldots & \mathbf{0} 
\end{bmatrix}^T$. Note that $\mathbf{c}_k$ is all zero for $k\geq M$.} of size $MS\times 1$ containing the $M$ first expectation values,
\begin{align}
\mathbf{A}
=
\begin{bmatrix}
\mathbf{A}_1 & \mathbf{A}_2 & \ldots & \mathbf{A}_{M-1} & \mathbf{A}_M\\
\mathbf{I} & \mathbf{0} & \ldots & \mathbf{0} & \mathbf{0}\\
\mathbf{0} & \mathbf{I} & \ldots & \mathbf{0} & \mathbf{0}\\
\vdots & \vdots & \ddots & \vdots & \vdots \\
\mathbf{0} & \mathbf{0} & \ldots & \mathbf{I} & \mathbf{0}\\
\end{bmatrix}
\label{eq:Amatrix}
\end{align}
is a block companion matrix of size $MS\times MS$,
\begin{align}
\mathbf{A}_m=
\begin{bmatrix}
a_{m11} & a_{m12} & \ldots & a_{m1S}\\
a_{m21} & a_{m22} & \ldots & a_{m2S}\\
\vdots & \vdots & \ddots &\vdots \\
a_{mS1} & a_{mS2} & \ldots & a_{mSS}
\end{bmatrix},
\label{eq:Ammatrix}
\end{align} 
is a size $S\times S$ submatrix, $m\in \{1,2,\ldots,M\}$, and
\begin{align}
a_{m\tilde{s}s}=\sum_{\nu=0}^{\nu_\emph{max}} P_{m \nu \tilde{s}s}\mathrm{e}^{-\theta R \nu}.
\label{eq:amss}
\end{align}
is a scalar entry for submatrix $\mathbf{A}_m$, for $\tilde s\in \{1,2,\ldots,S\}$, and $s\in \{1,2,\ldots,S\}$.
\end{theorem} 

\begin{IEEEproof}
The proof is structured as follows. We first establish that the state probability of transmission event state $\{k,n,s\}$ can be expressed as a linear combination of the state probabilities of the past transmission event states $\{k-m,n-\nu,\tilde{s}\}$, times the transition probabilities $P_{m \nu \tilde{s} s}$\footnote{This dependency may, at first, sound evident. What complicates things of HARQ (employing multiple transmissions), is that all potential sequences of transmission events do not terminate in state $\{k,n,s\}$, but some have longer transmission cycles going through $n$ and terminate just after $k$.}. Then, using this result, with some rearrangement of variables, we show that we can express the moment generating function (mgf) $\mathbb{E}\left\{ \mathrm{e}^{-\theta R N_k}\right \}$ through a system of recurrence relations. Following that, we rewrite the system of recurrence relations on a matrix recurrence form. 

We first rewrite the state probability as
\begin{align}
&\mathbb{P}\left\{ N_k=n,S_k=s \right \}\overset{(a)}{=} \sum_{\mu=0}^{M-1} d_{\mu  s} \pi_{k-\mu,n,s}\notag\\
&\overset{(b)}{=}  \sum_{\mu=0}^{M-1}  d_{\mu   s} \times \sum_{m=1}^M\sum_{\nu=0}^{\nu_\textrm{max}}  \sum_{\tilde{s}=1}^S P_{m\nu \tilde{s} s} \pi_{k-\mu-m,n-\nu,\tilde{s} }\notag\\
&\overset{(c)}{=}  \sum_{m=1}^{M} \sum_{\nu=0}^{\nu_\textrm{max}}  \sum_{\tilde{s}=1}^S P_{m\nu \tilde{s} s}  \times \sum_{\mu=0}^{M-1}   d_{\mu s} \pi_{(k-m)-\mu,n-\nu,\tilde{s}}\notag\\
&\overset{(d)}{=}\sum_{m=1}^M\sum_{\nu=0}^{\nu_\textrm{max}}  \sum_{\tilde{s}=1}^S  P_{m\nu \tilde{s} s} \mathbb{P}\left\{ N_{k-m}=n-\nu,S_k= \tilde{s} \right \},
\label{eq:Pstates}
\end{align}
where we have $d_{\mu  s}=\sum_{j=\mu+1}^{M} \sum_{\nu=0}^{\nu_\textrm{max}}  \sum_{s'=1}^S   P_{j \nu s s'}, \mu\in \{1,2,\ldots, M-1\}$ and $d_{0 s}=1$ in step (a). We have further used the dependency between current and past state termination probabilities, $\pi_{k,n,s}=\sum_{m=1}^M\sum_{\nu=0}^{\nu_\textrm{max}}  \sum_{\tilde{s}=1}^S P_{m\nu \tilde{s} s} \pi_{k-m,n-\nu,\tilde{s}}$, in step (b), changed the order of summation in step (c), and then used the equivalence of the first and second expression, as well as  assumed that $d_{\mu s}=d_{\mu \tilde{s}}, \forall \tilde{s}$, in step (d). The latter condition\footnote{This condition is not restrictive. One can also show that the recurrence relation framework can be developed based on the expression $\pi_{k,n,s}=\sum_{m=1}^M\sum_{\nu=0}^{\nu_\textrm{max}}  \sum_{\tilde{s}=1}^S P_{m\nu \tilde{s} s} \pi_{k-m,n-\nu,\tilde{s}}$ instead of using (\ref{eq:Pstates}), which then allows for $d_{\mu s}\neq d_{\mu \tilde{s}}, \forall \tilde{s}$, but the moment generating function must then be computed as a weighted sum of the $k$th recurrence relation vector.} means that all $\sum_{\nu=0}^{\nu_\textrm{max}}  \sum_{s'=1}^S   P_{m \nu s s'}$ are identical $\forall s$ when $m\geq2$, which is normally the case. We illustrate the above variables in Fig. \ref{fig:Transitions}.

We now show that $\mathbb{E}\left\{ \mathrm{e}^{-\theta R N_k}\right \}$ can be expanded into a system of recurrence relations which we arrive to in (\ref{eq:fkMatRec}). Considering the LHS of the recurrence first, we have
\begin{align}
\mathbb{E}\left\{ \mathrm{e}^{-\theta R N_k}\right \} 
= \sum_{s=1}^S  \sum_{\forall n} \mathrm{e}^{-\theta R n}\mathbb{P}\left\{ N_k=n,S_k=s\right \}.
\label{eq:LHS}
\end{align}
For the RHS of said recurrence relation, we get the expression
\begin{align}
&\mathbb{E}\left\{ \mathrm{e}^{-\theta R N_k}\right \}
= \sum_{s=1}^S  \sum_{\forall n} \mathrm{e}^{-\theta R n}\mathbb{P}\left\{ N_k=n,S_k=s\right \}\notag\\
& \overset{(a)}{=} \sum_{\forall n}\mathrm{e}^{-\theta R n}  \sum_{s=1}^S  \sum_{m=1}^M \sum_{\nu=0}^{\nu_\textrm{max}} \sum_{\tilde{s}=1}^S P_{m\nu \tilde{s} s} \notag\\
&\times \mathbb{P}\left\{ N_{k-m}=n-\nu,S_{k-m}=\tilde{s} \right \} \notag\\
&\overset{(b)}{=}\sum_{m=1}^M\sum_{\nu=0}^{\nu_\textrm{max}} \sum_{\tilde{s}=1}^S  \sum_{s=1}^S  P_{m\nu \tilde{s} s} \mathrm{e}^{-\theta R \nu}\notag\\
&\times \left( \sum_{\forall n}\mathrm{e}^{-\theta R (n-\nu)}\mathbb{P}\left\{ N_{k-m}=n-\nu,S_{k-m}=\tilde{s} \right \} \right),
\label{eq:RHS}
\end{align}
where we used (\ref{eq:Pstates}) in step (a), multiplied with $\mathrm{e}^{\theta R(\nu-\nu)}$, a dummy one, and changed the summation order in step (b). We now define $f_{k,s}\triangleq\sum_{\forall n} \mathrm{e}^{-\theta R n}\mathbb{P}\left\{ N_k=n,S_k=s\right \}$, and let
\begin{align}
 a_{m \tilde{s} s }  \triangleq \sum_{\nu=0}^{\nu_\textrm{max}} P_{m\nu \tilde{s} s} \mathrm{e}^{-\theta R \nu}.
\label{eq:a_m}
\end{align}
Using the above definitions, and equating the last expression in (\ref{eq:LHS}), and the last expression in (\ref{eq:RHS}), we find that
\begin{align}
\sum_{s=1}^S  f_{k,s}=  \sum_{s=1}^S  \sum_{\tilde{s}=1}^S  \sum_{m=1}^M  a_{m\tilde{s} s} f_{k-m,\tilde{s}},
\label{eq:RecScalar}
\end{align}
where $\mathbb{E}\left\{ \mathrm{e}^{-\theta R N_k}\right \}=\sum_{s=1}^S  f_{k,s}$. A structured strategy\footnote{Another strategy is to rewrite the (\ref{eq:RecScalar}) into a single homogeneous recurrence relation (with a resulting longer memory), by some strategic variable substitutions, and to solve the corresponding characteristic equation. Unfortunately, this is often tricky for a system of recurrence relations.} to solve (\ref{eq:RecScalar}) is to first reformulate it into a matrix recurrence relation, while omitting the summation over $s$. We then get the order-$m$ linear homogeneous matrix recurrence relation
\begin{align}
\mathbf{\tilde{f}}_{k}
=
\sum_{m=1}^M
\mathbf{A}_m 
\mathbf{\tilde{f}}_{k-m},
\label{eq:RecMatrixm}
\end{align}
where
\begin{align}
\mathbf{\tilde{f}}_{k}=
\begin{bmatrix} 
f_{k,1} & f_{k,2} & \ldots & f_{k,S} 
\end{bmatrix}^T,
\end{align}
and $\mathbf{A}_m$ as in (\ref{eq:Ammatrix}). Subsequently, (\ref{eq:RecMatrixm}) is  rewritten as a first order homogeneous linear matrix recurrence relation
\begin{align}
\mathbf{f}_{k}
=
\mathbf{A} 
\mathbf{f}_{k-1},
\label{eq:fkMatRec}
\end{align}
where
\begin{align}
\mathbf{f}_{k}=
\begin{bmatrix} 
\mathbf{\tilde{f}}_{k} & \mathbf{\tilde{f}}_{k-1} & \ldots & \mathbf{\tilde{f}}_{k-M} 
\end{bmatrix}^T,
\end{align}
and $\mathbf{A}$ is the block companion matrix in (\ref{eq:Amatrix}).

We can now finalize the proof of the effective capacity expression by noting that
\begin{align}
C_{\textrm{eff},k}^\textrm{Retr.}
&=-\frac{1}{\theta k} \ln{\left(\mathbb{E}\{\mathrm{e}^{-\theta R N_k}\}\right)}\notag\\
&=-\frac{1}{\theta k} \ln{\left(\sum_{s=1}^S f_{k,s}\right)}\notag\\
&\overset{(a)}{=}-\frac{1}{\theta k} \ln{\left(\mathbf{b}^T\mathbf{f}_k\right)}\notag\\
& \overset{(b)}{=}-\frac{1}{\theta k} \ln{\left(\mathbf{b}^T\mathbf{A}^{k-M} \mathbf{f}_{M}\right)},\ k\geq M,
\label{eq:CeffRetrSchemeProof}
\end{align}
where the sum is put on a vector-matrix form in step (a), and we use  $\mathbf{f}_k=\mathbf{A} \mathbf{f}_{k-1}$ repeatedly in step (b).
\end{IEEEproof}

Eq. (\ref{eq:CeffRetrk}) reveals that the effective capacity for a $k$-timeslot retransmissions system is fully determined by the block companion matrix $\mathbf{A}$, and the initiation vector $\mathbf{f}_0$. We now see that, in contrast to related work on effective capacity, \cite{bibChang}, \cite{bibWu}-\cite{bibAkin2}, that consider the limiting case $k\rightarrow \infty$, Theorem \ref{theorem:CeffRetrScheme} gives an algebraic matrix expression of the log-mgf (and the effective capacity) for any $k$.
An important aspect to note from Theorem \ref{theorem:CeffRetrScheme}, is that theorem enable us to also compute the ($k$-timeslot) throughput. This is so since
\begin{align}
\lim_{\theta\rightarrow 0}C_{\textrm{eff},k}^\textrm{Retr.}
=\frac{R\mathbb{E}\{N_k\}}{k}
\triangleq T_k^\textrm{Retr.},
\label{eq:CkTk}
\end{align}
which is proved in Appendix \ref{sec:CeffConverT}.
In Appendix \ref{sec:kTSThroughput}, we also show that the idea of a recurrence relation formulation, as used in Theorem \ref{theorem:CeffRetrScheme}, can be used to analyze the $\alpha$-moment, $\mathbb{E}\{N_k^\alpha\}$. This in turn, allows the throughput (using the first-moment $\mathbb{E}\{N_k\}$), to be determined directly rather than as a limit in (\ref{eq:CkTk}). From now on, we will focus on the limiting case, $k \rightarrow \infty$, of the effective capacity, (\ref{eq:Ceff}). With this in mind, we first give the effective capacity for infinite $k$ in the Corollary below.
\begin{Corollary}
\label{Corollary:CeffRetrSchemekInf}
The effective capacity of a retransmission scheme, with $k \rightarrow  \infty$, $M$ as transmission limit, $S$ communication modes, $\nu\in\{0,1,\ldots,\nu_\emph{max}\}$ packets per transmission cycle, and the transition probabilities $P_{m \nu \tilde{s} s}$, is 
\begin{align}
C_\textrm{eff}^\textrm{Retr.}
&=-\frac{\ln{(\lambda_+)}}{\theta},
\label{eq:Cefflambda}
\end{align} 
where $\lambda_+=\max\{|\lambda_1|,|\lambda_2|,\ldots,|\lambda_{MS}|\}$ is the spectral radius of the block companion matrix $\mathbf{A}$, with eigenvalues $\{\lambda_1,\lambda_2,\ldots,\lambda_{MS}\}$, given in Theorem \ref{theorem:CeffRetrScheme}.
\end{Corollary} 
\begin{IEEEproof}
From Theorem \ref{theorem:CeffRetrScheme}, we get the expression
\begin{align}
C_\textrm{eff}^\textrm{Retr.}
&=-\lim_{k\rightarrow \infty}\frac{1}{\theta k}\ln{\left(\mathbf{b}^T\mathbf{A}^{k-M} \mathbf{f}_{M}\right)} \notag\\
&\overset{(a)}{=}- \lim_{k\rightarrow \infty}\frac{1}{\theta k} \ln{\left(\mathbf{b}^T\mathbf{Q}\mathbf{\Lambda}^{k-M}\mathbf{Q}^{-1} \mathbf{f}_{M}\right)}\notag\\
&\overset{(b)}{=}- \lim_{k\rightarrow \infty}\frac{1}{\theta k} \ln{\left(\lambda_+^{k-M}\mathbf{b}^T\mathbf{Q}(\mathbf{\Lambda}/\lambda_+)^{k-M}\mathbf{Q}^{-1} \mathbf{f}_{M}\right)}\notag\\
& \overset{(c)}{=}-\frac{\ln{\left(\lambda_+\right)}}{\theta},
\label{eq:Ceffprooflimit}
\end{align}
where the eigendecomposition $\mathbf{A}=\mathbf{Q}\mathbf{\Lambda}\mathbf{Q}^{-1}$ is exploited in step (a), the largest absolute eigenvalues of $\mathbf{A}$, $\lambda_+$, is expanded for in step (b), and $\lambda_+$ is shown to dominate as $k\rightarrow \infty$ in step (c). We note that Corollary \ref{Corollary:CeffRetrSchemekInf} holds even if the transmission limit $M$ increases with time $k$, as long as $M$ increases slower than linearly with $k$.
\end{IEEEproof}

To illustrate the usefulness of the unified approach, Theorem \ref{theorem:CeffRetrScheme} and Corollary \ref{Corollary:CeffRetrSchemekInf}, we now consider and analyze practically interesting schemes, such as truncated-HARQ in Section \ref{sec:HARQ}, and NC-ARQ in Section \ref{sec:NCARQ}. In Sections \ref{sec:HARQwireless} and \ref{sec:HARQ_MED}, we leave the (finite size) matrix formulation in Theorem 1 and Corollary 1, which is also the basis for \cite{bibTang2}-\cite{bibAkin2}, and instead consider a characteristic equation form allowing for an infinite transmission limit $M$.

\subsection{HARQ}
\label{sec:HARQ}
In this section, we focus on truncated-HARQ where the probabilities, $P_m, m\in \{1,2,\ldots,M\}$ and $Q_M$, are assumed given. For this case, we have only one communication mode, $S=1$, and a packet of rate $R$ is either delivered at latest on the $M$th transmit attempt, or is not delivered at all. The Corollary below enables the corresponding effective capacity to be computed.
\begin{Corollary}
\label{Corollary:CeffHARQ}
The effective capacity of truncated-HARQ, with transmission limit $M$, $\nu\in\{0,1\}$ packet per transmission cycle, and the probabilities of successful decoding on the $m$th attempt $P_m$, $Q_M\triangleq 1-\sum_{m=1}^M P_m$, is given by Theorem \ref{theorem:CeffRetrScheme} for finite $k$ (or Corollary \ref{Corollary:CeffRetrSchemekInf} for infinite $k$), together with the companion matrix
\begin{align}
\mathbf{A}
=
\begin{bmatrix}
a_1 & a_2 & \ldots & a_{M-1} & a_M\\
1 & 0 & \ldots & 0 & 0\\
0 & 1 & \ldots & 0 & 0\\
\vdots & \vdots & \ddots & \vdots & \vdots \\
0 & 0 & \ldots & 1 & 0\\
\end{bmatrix},
\label{eq:AHARQmatrix}
\end{align}
where
\begin{align}
 a_{m} \triangleq &
    \begin{cases}
       P_{m}\mathrm{e}^{-\theta R }, & m\in \{1,2,\ldots M-1\}, \\
      P_{M}\mathrm{e}^{-\theta R } +Q_M, & m=M. \\
    \end{cases}
\label{eq:HARQam}
\end{align}
\end{Corollary} 
\begin{IEEEproof}
Corollary \ref{Corollary:CeffHARQ} follows directly from Theorem \ref{theorem:CeffRetrScheme} with $S=1$, with $P_m\neq 0, m\in\{1,2,\ldots,M\}$ and $Q_M\neq 0$, where we introduced and defined the decoding success probabilities $P_m\triangleq P_{m111}$ and the decoding failure probability $Q_M\triangleq P_{M011}$. Since Corollary \ref{Corollary:CeffRetrSchemekInf} followed from Theorem \ref{theorem:CeffRetrScheme}, it uses the spectral radius of (\ref{eq:AHARQmatrix}) with entries (\ref{eq:HARQam}).
\end{IEEEproof}
While a full derivation for a general retransmission scheme is performed in Theorem  \ref{theorem:CeffRetrScheme}, it is nevertheless instructive and useful to highlight some of the key-expressions for truncated-HARQ. Simplifying the notation from Theorem \ref{theorem:CeffRetrScheme}, with $f_k\triangleq f_{k,1}$ and $a_m\triangleq a_{m11}$, we see that the recurrence relation (\ref{eq:RecScalar}), for one communication mode, can be written as the homogeneous recurrence relation
\begin{align}
f_k=\sum_{m=1}^M a_m f_{k-m},
\label{eq:f_kHARQ}
\end{align}
where $f_k=\mathbb{E}\{\mathrm{e}^{-\theta R N_k}\}$. Alternatively, (\ref{eq:f_kHARQ}) can be written on the matrix recurrence relation form 
\begin{align}
\mathbf{f}_k=\mathbf{A}\mathbf{f}_{k-1},
\end{align}
where $\mathbf{f}_k=[f_{k}\ f_{k-1}\  \ldots \ f_{k-M}]^T$, and $\mathbf{A}$ is given by (\ref{eq:AHARQmatrix}). An alternative to the spectral radius of (\ref{eq:AHARQmatrix}), is to determine the largest root of the characteristic equation of (\ref{eq:f_kHARQ}). Setting $f_k=\lambda^k$ in (\ref{eq:f_kHARQ}), the characteristic equation is found to be
\begin{align}
\lambda^M-\sum_{m=1}^M a_m\lambda^{M-m}=0.
\label{eq:HARQCharEq}
\end{align}
This formulation, (\ref{eq:HARQCharEq}), is (as will be seen) vital for validation purpose and extending the analysis to persistent-HARQ, in fact the basis for Corollary \ref{Corollary:Valid1}-\ref{Corollary:CpsiRRRayleigh}. For convenience to the reader, we also show the simpler, and more tractable, derivation of the recurrence relation for truncated-HARQ in Appendix \ref{ECHARQApp}. 

Here it can be noted that matrix $\mathbf{A}$, (\ref{eq:AHARQmatrix}), with entries (\ref{eq:HARQam}), is not on the $\mathbf{P}\boldsymbol{\Phi}$-form (which is due to the Markov modulated process modeling) as in \cite{bibTang2}-\cite{bibAkin2}. It is easy to see that this also holds true more generally for (\ref{eq:Amatrix}) with (\ref{eq:Ammatrix}). Observe also that Corollary \ref{Corollary:CeffHARQ}, with the expressions (\ref{eq:AHARQmatrix}) and (\ref{eq:HARQam}) differs from \cite[(9)]{bibAkin2}. This is so for two important reasons. The first reason is that mathematical model in the latter implies that the packet is always delivered before or at the last transmit attempt, whereas in this work, a data packet is discarded on the last transmit attempt if decoding fails. The second reason is that the Markov modulated process modeling gives a more complicated matrix-form, with e.g. ratios of transition probabilities and transition probabilities repeated in multiple entries. The RW and recurrence relation framework, not only give (\ref{eq:AHARQmatrix}) on a simple form, but also (\ref{eq:HARQCharEq}) which enables the derivation of Corollary \ref{Corollary:Valid1}-\ref{Corollary:CpsiRR}.

We now continue with three subsections. The first two give validation and application examples, whereas the third frame the effective capacity more directly in real-world QoS-parameters, i.e. the delay target $D_\textrm{max}$ and the delay violation probability $\epsilon$, rather than in the QoS-exponent $\theta$.

\subsubsection{Validation Examples}
\label{sec:Validation}
This section serves to validate the soundness of the HARQ analysis above. A first aspect to investigate is that the effective capacity is in the range $(0,\infty)$, which is done in the following Corollary.
\label{sec:ValidationEx}
\begin{Corollary}
\label{Corollary:Valid1}
The characteristic equation, $\lambda^M-\sum_{m=1}^M a_m\lambda^{M-m}=0$, $a_m>0$, has only one positive root, and it lies in the interval $[0,1]$.  
\end{Corollary} 
\begin{IEEEproof}
The proof is given in Appendix \ref{sec:ProofValid1}\footnote{We note that the same goal as for Corollary \ref{Corollary:Valid1} have also been considered in \cite[Thm2]{bibAkin2}. However, Corollary \ref{Corollary:Valid1} with proof differs. First, the characteristic equation differs from \cite[(4)]{bibAkin2} since in the present analytical model, a packet may be discarded when reaching the transmission limit. Second, a slightly different approach, exploiting Descarte's rule of signs, is used to show that only one positive root exists. Third, but not considered in \cite[Thm2]{bibAkin2}, it is shown that the positive root lies in the interval [0,1] which is required for a real and positive effective capacity.}.
\end{IEEEproof} 
Thus, this confirms that $0\leq C_\textrm{eff}^\textrm{HARQ}\leq \infty$, since $0 \leq \lambda_+ \leq 1$, and $C_\textrm{eff}^\textrm{HARQ}=-\ln(\lambda_+)/\theta$.

The next Corollary verifies that the effective capacity, Corollary \ref{Corollary:CeffRetrSchemekInf} with (\ref{eq:AHARQmatrix}) and (\ref{eq:HARQam}), converges to the well-known throughput expression (\ref{eq:THARQtrunc}) of truncated-HARQ. 

\begin{Corollary}
\label{Corollary:ECtoTHARQ}
The effective capacity of truncated-HARQ converges to the throughput of  truncated-HARQ as $\theta\rightarrow 0$,
\begin{align}
\lim_{\theta\rightarrow 0}C_\textrm{eff}^\textrm{HARQ}=\frac{R(1-Q_M)}{\sum_{m=1}^M m P_m+MQ_M }\triangleq T^\textrm{HARQ}_\textrm{trunc.}.
\label{eq:CeffTHARQ}
\end{align} 
\end{Corollary} 

\begin{IEEEproof}
The proof is given in Appendix \ref{sec:ProofValid2}.
\end{IEEEproof} 
Note that when $M\rightarrow \infty$, i.e. $Q_M\rightarrow 0$,  (\ref{eq:CeffTHARQ}) converges to the throughput of persistent-HARQ in (\ref{eq:THARQloss}). For the case with $\theta\rightarrow \infty$, the characteristic equation yields $\lambda_+=Q_M^{1/M}$, which gives $ \lim_{\theta\rightarrow \infty}C_\textrm{eff}^\textrm{HARQ}=-\lim_{\theta\rightarrow \infty}\frac{\ln{(Q_M)}}{\theta M}=0$. 

To further validate Corollary \ref{Corollary:CeffHARQ}, and the forms of (\ref{eq:AHARQmatrix}), (\ref{eq:HARQam}), we now show that the effective capacity of truncated-HARQ  degenerates to the effective capacity of ARQ (\ref{CeffARQ}) when $P_m$ is geometrically distributed. This need to be the case since this, the assumed transmission independence, implies that information from earlier transmissions is not exploited.

\begin{Corollary}
\label{Corollary:ECGeomARQ}
The effective capacity for truncated-HARQ with $P_m$ geometrically distributed, $P_m=P_1Q_1^{m-1}$, $P_1=1-Q_1$,  $m\in  \{1,2,\ldots ,M \}$, and $Q_M=1-\sum_{m=1}^M P_m=Q_1^M$ is
\begin{align}
C_\textrm{eff}^\textrm{HARQGeom}=-\frac{1}{\theta}\ln {\left( Q_1 +P_1\mathrm{e}^{-\theta R}\right) }.
\label{eq:CeffGeom}
\end{align} 
\end{Corollary}  

\begin{IEEEproof}
The proof is given in Appendix \ref{sec:ProofValid3}.
\end{IEEEproof}
As expected, we note that (\ref{eq:CeffGeom}) has exactly the same form as the effective capacity of ARQ (\ref{CeffARQ}). We now turn the attention to two application examples of the truncated-HARQ analysis, specifically of (\ref{eq:HARQCharEq}).

\subsubsection{Application Examples}
\label{sec:ApplicationEx}
Here, we apply the truncated-HARQ analysis to the simplest truncated-HARQ system imaginable, i.e. with a maximum of $M=2$ transmissions\footnote{Another interesting case is for $M\rightarrow \infty$, but this requires some more assumptions and is therefore handled in Section \ref{sec:HARQwireless}.}, and also give an approximative effective capacity expression that is valid for small $\theta$.

We start with this simple truncated-HARQ case, also shown in Fig. \ref{fig:Transitions}. The following Corollary gives a closed-form expression for the effective capacity of such system.
\begin{Corollary}
\label{Corollary:Ceff2}
The effective capacity for truncated-HARQ with transmission limit $M=2$, and probabilities $P_1$, $P_2$ of successful decoding, and probability $Q_2=1-P_1-P_2$ of failed decoding, is
\begin{align}
C_\emph{eff}^\emph{HARQ}=R-\frac{1}{\theta}\ln\left\{ \frac{P_1+\sqrt{P_1^2+4(P_2\mathrm{e}^{\theta R}+Q_2\mathrm{e}^{\theta 2R})}}{2} \right\}.
\label{eq:Ceff2}
\end{align} 
\end{Corollary} 

\begin{IEEEproof}
The recurrence relation for this system is $f_k=a_1f_{k-1}+a_2f_{k-2}$. This is a
a recurrence relation for a bivariate Fibonacci-like polynomial\footnote{The Fibonacci numbers is given with $a_1=a_2=1$, $f_0=0$, and $f_1=1$.} \cite[pp. 152-154]{bibHazewinkel}, with $a_1=P_1\mathrm{e}^{-\theta R}$, $a_2=P_2\mathrm{e}^{-\theta R}+Q_2$. The characteristic equation is $\lambda^2-a_1\lambda-a_2=0$,
with the largest positive root
\begin{align}
\lambda_+=\frac{1}{2}\left(a_1+\sqrt{a_1^2+4a_2}\right).
\label{eq:Fibalpha+}
\end{align}
Inserting (\ref{eq:Fibalpha+}) in (\ref{eq:Cefflambda}), and then bringing out $R$, gives (\ref{eq:Ceff2}).
\end{IEEEproof}.

We now turn our attention to the approximation of the effective capacity in terms of the first and second moments of the number of transmissions per packet.
\begin{Corollary}
\label{Corollary:CeffAppr}
The effective capacity for persistent-HARQ can, for small $\theta$, be approximated as
\begin{align}
C_\emph{eff}^\emph{HARQ}&\approx \frac{c_1+\sqrt{c_1^2+4c_2}}{2},\\
\label{eq:CeffAppr}
c_1&=\frac{2\mu(1-\theta R) }{\theta(\sigma^2+\mu^2)}, \,
c_2=\frac{ R(2-\theta R)}{\theta(\sigma^2+\mu^2)},
\end{align} 
 where $\mu=\sum_{m=1}^\infty mP_m$ signify the mean, and $\sigma^2=\sum_{m=1}^\infty m^2P_m$ denotes the variance of the number of transmissions per packet.
\end{Corollary} 

\begin{IEEEproof}
The proof is given in Appendix \ref{sec:ProofAppl2}.
\end{IEEEproof} 
We note that Corollary \ref{Corollary:CeffAppr} offers a new effective capacity approximation of HARQ in the first- and second-order moments\footnote{An approximation for truncated-HARQ can also be found, but is then, in addition to $\mu$ and $\sigma$, also expressed with $Q_M\neq 0$.}. This is similar in spirit to the approximation $C_\textrm{eff}^\textrm{HARQ}\approx R/\mu-R^2\sigma^2\theta/2\mu^3$, for $\theta \approx 0$,  proposed in \cite{bibLi}.

\subsubsection{Effective Capacity Expressed in QoS-Parameters}
\label{sec:ECQoS}
So far, we have considered the effective capacity for a given $\theta$. Taking on, more of, an engineering point of view, we are also interested in its dependency on $D_\textrm{max}$ and $\epsilon$. This is considered in the Corollary below. The main idea is to first express $\theta$ as a function of $\theta C_\textrm{eff}$, then from (\ref{eq:thetaC}) to use the substitution $\psi=\theta C_\textrm{eff}$  in the expression for $\theta$, and lastly to rearrange (\ref{eq:thetaC}) into $C_\textrm{eff}(\psi)=\psi/\theta(\psi)$.
\begin{Corollary}
\label{Corollary:CpsiHARQ1} 
The effective capacity of truncated-HARQ, with transmission limit $M$, $\nu\in\{0,1\}$, probabilities of successful decoding on the $m$th attempt $P_m$, $Q_M\triangleq 1-\sum_{m=1}^M P_m$,  and $\psi\triangleq \log(\eta/\epsilon)/D_\textrm{max}$, is 
\begin{align}
C_\emph{eff}^\emph{HARQ}(\psi)
&=\frac{R\psi}{\ln \left(\sum_{m=1}^MP_m\mathrm{e}^{\psi m} \right)-\ln\left(1-Q_M\mathrm{e}^{\psi M}\right)} 
\label{eq:CpsiHARQ1}
\end{align}
\end{Corollary}
\begin{IEEEproof}
We rewrite (\ref{eq:thetaC}) as $C_\textrm{eff}(\psi)=\psi/\theta(\psi)$, and then rewrite (\ref{eq:HARQCharEq}) as
$\mathrm{e}^{\theta R}=(1-Q_M\lambda^{-M})^{-1}\sum_{m=1}^M P_m \lambda^{-m}$, which is solved for $\theta$. Inserting $\theta$, with $\lambda\triangleq \mathrm{e}^{-\theta C_\textrm{eff}}=\mathrm{e}^{-\psi}$, in  $C_\textrm{eff}(\psi)=\psi/\theta(\psi)$ concludes the proof.
\end{IEEEproof}
We note that (\ref{eq:CpsiHARQ1}) reduces to the classical throughput expression (\ref{eq:THARQtrunc})  if $\psi\rightarrow 0$, and that persistent-HARQ has the simple and compact expression $C_\textrm{eff}^\textrm{HARQ}(\psi)
=R\psi/\ln \left(\sum_{m=1}^\infty P_m\mathrm{e}^{\psi m} \right)$. Observe that this alternative expression is a generalization of the well-known throughput expression (\ref{eq:THARQloss}). Note also that $\psi \leq -\ln{(Q_M)}/M$ for a real denominator in (\ref{eq:CpsiHARQ1}).
Another observation is that one can plot (\ref{eq:CpsiHARQ1}) vs, $\theta$ parametrically as $(\psi/C_\textrm{eff}(\psi),C_\textrm{eff}(\psi))$. Eq. (\ref{eq:CpsiHARQ1}) is also interesting since it suggest, and we conjecture, that $C_\textrm{eff}(\psi)\triangleq\lim_{n\rightarrow \infty}{R\psi}/{\ln\left(\mathbb{E}\{\mathrm{e}^{\psi K_n}\} \right)}$, where  $K_n$ is a r.v. for the number of timeslots used to deliver $n$ packets. Applying the RW idea, and letting $f_n\triangleq \mathbb{E}\{\mathrm{e}^{\psi K_n}\}$, we can formulate the recurrence $f_n=\left(\sum_{m=1}^M P_m \mathrm{e}^{m\psi}\right)f_{n-1}+Q_M \mathrm{e}^{M\psi}f_{n}$. This recurrence has the characteristic equation solution $\lambda_+=(1-Q_M \mathrm{e}^{M\psi})^{-1}(\sum_{m=1}^M P_m \mathrm{e}^{m\psi})$, which agrees with (\ref{eq:CpsiHARQ1}).

\subsection{HARQ with Effective Channel}
\label{sec:HARQwireless}
In Theorem \ref{theorem:CeffRetrScheme}, we gave an effective capacity expression for general retransmission schemes in terms of transition probabilities. In Section \ref{sec:HARQ}, we specialized this result to HARQ. Now, this specialized expression is used to derive an effective capacity expression where the effective channel is described by a pdf $f_Z(z)$, its Laplace transform $F(s)$, and a decoding threshold $\Theta$. A second fundamental result of the paper is given by the following theorem.
\begin{theorem}
\label{theorem:ECHARQEffCh}
For persistent-HARQ schemes, characterized by an effective channel pdf $f_Z(z)$ and threshold $\Theta$, the spectral radius, $\lambda_+$, in (\ref{eq:Cefflambda}), is implicitly given by
\begin{align}
\mathrm{e}^{\theta R}&= \mathcal{L}_\Theta^{-1} \left \{ \frac{1}{s}\frac{1-F(s)}{\lambda_+-F(s)} \right \}
\label{eq:alphaimplicitpersistent}
\end{align}
\end{theorem}

\begin{IEEEproof}
Divide (\ref{eq:HARQCharEq}) by $\lambda^M$, and then let $k\rightarrow \infty$. This allows for $M \rightarrow \infty$, as long as $M$ increases less than linearly with $k$, and the resulting characteristic equation for persistent-HARQ becomes
\begin{align}
1=\sum_{m=1}^\infty P_m\mathrm{e}^{-\theta R} \lambda^{-m}.
\end{align}
We rewrite this characteristic equation as 
\begin{align}
 &\mathrm{e}^{\theta R}\overset{(a)}{=}\sum_{m=1}^\infty (Q_{m-1}-Q_m) \lambda^{-m}\notag\\
&\overset{(b)}{=}\sum_{m=1}^\infty \left(\int_0^\Theta \!\!\! \mathcal{L}_z ^{-1} \left \{ F(s)^{m-1} -F(s)^m \right \} \! \mathrm{d}z \right) \lambda^{-m}\notag\\
&\overset{(c)}{=} \int_0^\Theta \!\!\! \mathcal{L}_z^{-1} \left \{\sum_{m=1}^\infty \frac{1}{F(s)}\left(\frac{F(s)}{\lambda}\right )^{m} -\left(\frac{F(s)}{\lambda}\right)^m \right \}  \! \mathrm{d}z \notag\\
&\overset{(d)}{= }\int_0^\Theta \mathcal{L}_z^{-1} \left \{ \frac{1-F(s)}{\lambda-F(s)} \right \} \, \mathrm{d}z
\overset{(e)}{=} \mathcal{L}_\Theta ^{-1}\left \{ \frac{1}{s}\frac{1-F(s)}{\lambda-F(s)} \right \},
\end{align}
where we used $P_m=Q_{m-1}-Q_m$ in step (a), (\ref{eq:Qm}) in step (b), changed the sum and the integration order in step (c), computed the geometric series in step (d), and applied the Laplace transform integration rule in step (e).
\end{IEEEproof}
\begin{remark}
It is, in principle, possible to extend and generalize the idea behind Theorem \ref{theorem:ECHARQEffCh} to the wider scope of Theorem \ref{theorem:CeffRetrScheme}. This could, e.g., allow $S\geq2$, with multiple effective channels, to be handled as $M\rightarrow\infty$. However, this goes somewhat beyond the scope of the paper, and is omitted in the following.
\end{remark}
A similar derivation for truncated-HARQ is possible, but yields the (somewhat less appealing) form
\begin{align}
\mathrm{e}^{\theta R}
&=\left( \mathcal{L}_\Theta^{-1} \left \{ \frac{1}{s} \left(1- \frac{F(s)^M}{\lambda_+^M} \right)  \right \} \right)^{-1}\notag\\
&\times\mathcal{L}_\Theta^{-1} \left \{ \frac{1}{s} 
\frac{1-F(s)}{\lambda_+-F(s)}\left(1+\frac{F(s)^M}{\lambda_+^M} \right) \right \}.
\end{align}

Also for the effective channel case, it is of interest to express $C_\textrm{eff}$ in the QOS-parameter $\psi$. This is done in Corollary \ref{Corollary:CpsiRR} where we focus on the persistent-HARQ case.
\begin{Corollary} 
\label{Corollary:CpsiRR}
The effective capacity of persistent-HARQ, characterized by an effective channel pdf $f_Z(z)$, threshold $\Theta$, and $\psi\triangleq \log(\eta/\epsilon)/D_\textrm{max}$, is 
\begin{align}
C_\textrm{eff}^\textrm{HARQ}(\psi)
=\frac{R}{\psi^{-1}\ln\left( \mathcal{L}_\Theta^{-1}\left\{\frac{\mathrm{e}^\psi}{s}\frac{1-F(s)}{1-F(s)\mathrm{e}^\psi}\right\}\right)}.
\label{eq:CeffHARQPsi}
\end{align}
\end{Corollary}
\begin{IEEEproof}
We use Theorem \ref{theorem:ECHARQEffCh} with $\lambda=\mathrm{e}^{-\psi}$, $C(\psi)=\psi/\theta(\psi)$, and then rearrange the expression. 
\end{IEEEproof}
Thus, with $F(s)$ given, we can either use (\ref{eq:alphaimplicitpersistent}) and solve for $\lambda_+$ to compute the effective capacity in terms of $\theta$, or we can use (\ref{eq:CeffHARQPsi}), to get the effective capacity in terms of $\psi$. The benefit of the latter is the closed-form expression and relating the effective capacity directly to delay target and delay violation probability. We now turn our attention to the lowest, the third, system model level and consider fading channel models.

\subsection{HARQ with ME- / Rayleigh-Distributed Fading Channels}
\label{sec:HARQ_MED}
We observe that Theorem \ref{theorem:ECHARQEffCh} and Corollary \ref{Corollary:CpsiRR}, expressed in $F(s)$, makes them amenable to integrate with the compact and powerful MED effective channel framework for wireless channels introduced in \cite{bibLarsson2}, and also reviewed in Section \ref{sec:SystModel3}. We start by studying the persistent-HARQ case in the following Corollary.
\begin{Corollary} 
\label{Corollary:CpsiMEDRR}
The effective capacity of persistent-HARQ, characterized by an effective channel MED pdf $f_Z(s)=\mathbf{p}\mathrm{e}^{z\mathbf{Q}}\mathbf{r}$, $\mathbf{Q}=\mathbf{S}-\mathbf{r}\mathbf{q}$, threshold $\Theta$, and $\psi\triangleq \log(\eta/\epsilon)/D_\textrm{max}$, is 
\begin{align}
C_\textrm{eff}^\textrm{HARQ}(\psi)
=\frac{R}{\psi^{-1}\ln\left( \mathbf{a}\mathrm{e}^{\Theta\mathbf{B}}\mathbf{c}\right)},
\label{eq:RRCeffRpsi}
\end{align}
where $\mathbf{a}=[0 \ (\mathbf{q}-\mathbf{p})\mathrm{e}^\psi]$, $\mathbf{B}=\mathbf{S}-\mathbf{c}[0 \ (\mathbf{q}-\mathbf{p}\mathrm{e}^\psi)]$, $\mathbf{c}=[0; \mathbf{r}]$.
\end{Corollary}
\begin{IEEEproof}
Using Corollary \ref{Corollary:CpsiRR}, with $F(s)=p(s)/q(s)$, gives the argument $\mathcal{L}^{-1}_\Theta\left\{\frac{\mathrm{e}^\psi}{s}\frac{q(s)-p(s)}{q(s)-p(s)\mathrm{e}^\psi}\right\}$ of the logarithm. We then let $a(s)=\mathrm{e}^\psi(q(s)-p(s))$ and $b(s)=s(q(s)-p(s)\mathrm{e}^\psi)$ and insert the coefficients in the MED-form.
\end{IEEEproof}
Many works in the literature consider throughput optimization for (H)ARQ. Likewise, it is of interest to find the maximum effective capacity of (\ref{eq:RRCeffRpsi}) wrt the rate $R$, and the optimal rate point $R^*$. The classical optimization approach is to consider $\mathrm{d}C_\textrm{eff}/\mathrm{d}R=0$, and solve for $R^*(\Gamma)$. This approach is (generally) not possible here, since a closed-form expression for the optimal rate-point is hard (or impossible) to find. We therefore resort to the auxiliary parametric optimization method (method 2) that we developed in \cite{bibLarsson1}, and show below that it also handles effective capacity optimization problems.
\begin{Corollary} 
\label{Corollary:CpsiMEDRRopt}
The optimal effective capacity of persistent-HARQ, characterized by an effective channel MED pdf $f_Z(s)=\mathbf{p}\mathrm{e}^{z\mathbf{Q}}\mathbf{r}$, $\mathbf{Q}=\mathbf{S}-\mathbf{r}\mathbf{q}$, threshold $\Theta$, and $\psi\triangleq \log(\eta/\epsilon)/D_\textrm{max}$, is 
\begin{align}
f_\Theta(\psi,\Theta)
&\triangleq 
\frac{\mathbf{a}\mathrm{e}^{\Theta\mathbf{B}}\mathbf{c}}
{\mathbf{a}\mathrm{e}^{\Theta\mathbf{B}}\Theta\mathbf{B}\mathbf{c}}\ln\left( \mathbf{a}\mathrm{e}^{\Theta\mathbf{B}}\mathbf{c}\right), \\
\label{eq:RRoptCeffRpsifTheta}
R^*(\psi,\Theta)&=\lg_2(\mathrm{e})\left(f_\Theta+W_0(-f_\Theta\mathrm{e}^{-f_\Theta})\right),\\
\Gamma(\psi,\Theta)&=\frac{2^{R^*}-1}{\Theta},\\
C_\textrm{eff}^\textrm{HARQ*}(\psi,\Theta)
&=\frac{R^*}{\psi^{-1}\ln\left( \mathbf{a}\mathrm{e}^{\Theta\mathbf{B}}\mathbf{c}\right)},
\label{eq:RRoptCeffRpsi}
\end{align}
where $0\leq\Theta<\infty$ is the auxiliary parameter, and $f_\Theta(\psi,\Theta)$ is a function facilitating more compact expressions.
\end{Corollary}
\begin{IEEEproof}
Assuming a (local) maxima exists, we take the derivative of the log of (\ref{eq:RRCeffRpsi}), noting that $\Theta$ is a function of $R$, equate it to zero, and arrange $R^*$- and $\Theta$-dependent terms on the LHS and RHS of the equal sign. The $\theta$-dependent function on the RHS is denoted by $f_\Theta$, and the LHS is solved for $R^*$. We refer the interested reader to \cite{bibLarsson1} for more details.  
\end{IEEEproof}
\begin{figure}[t]
 \centering
 \includegraphics[width=9cm]{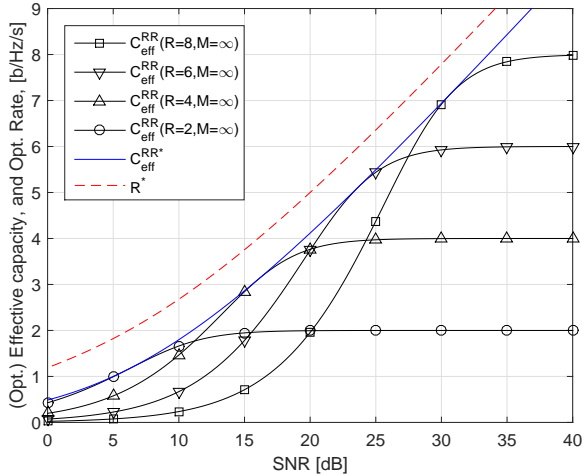}
 \vspace{-0.75cm}
 \caption{The effective capacity vs. SNR for persistent-RR with $N=2$ (Alamouti-Tarokh's TX-diversity), Rayleigh fading, $\psi=1$, for $R=\{2,4,6,8\}$ b/Hz/s (\ref{eq:RRCeffRpsi}). The optimal effective capacity and the optimal rate point vs. SNR for the same scenario (\ref{eq:RRoptCeffRpsifTheta})-(\ref{eq:RRoptCeffRpsi}).}
 \label{fig:CeffRROpt}
 \vspace{-0.25cm}
\end{figure}
In Fig. \ref{fig:CeffRROpt}, we plot the effective capacity expression (\ref{eq:RRCeffRpsi}) vs. SNR for four different rates, Rayleigh fading and Alamouti-transmit diversity. We first note that the MED-channel model works fine in conjunction with the effective capacity metric. The characteristics shown in Fig. \ref{fig:CeffRROpt} are as expected, i.e. the effective capacity increases from 0 at $\Gamma=-\infty$ dB to $R$ at $\Gamma=\infty$ dB.  We can also verify that (\ref{eq:RRoptCeffRpsifTheta})-(\ref{eq:RRoptCeffRpsi}), allows the maximum effective capacity and the optimal rate point to be plotted.
Corollary \ref{Corollary:CpsiMEDRRopt} handles the optimization of the effective capacity expressed in $\psi$ and the MED-channel. 
Due the implicit structure of (\ref{eq:alphaimplicitpersistent}), i.e. the need of solving for  $\lambda_+$, it is often hard to solve (\ref{eq:alphaimplicitpersistent}), even for the MED-based effective channel case. Nevertheless, in the following Corollary, we show how to solve  (\ref{eq:alphaimplicitpersistent}) for RR operating in a Rayleigh fading channel. This also applies to the OSTBC-MRC-Nakagami-$m$ effective channel in Section \ref{sec:SystModel3}.
\begin{Corollary}
\label{Corollary:SISORR}
The effective capacity of RR in block Rayleigh fading, with $\tilde \Theta =(2^R-1)/\Gamma$, is
\begin{align}
C_\textrm{eff}^\emph{RR}=R-\frac{1}{\theta}\left( W_0 \left( \tilde\Theta \mathrm{e}^{\tilde\Theta+\theta R}\right) -\tilde\Theta \right),
\label{eq:SISORRtheta}
\end{align}
and the spectral radius, $ \lambda_+$, is
\begin{align}
\lambda_+=\frac{\tilde\Theta}{W_0 \left( \tilde\Theta \mathrm{e}^{\tilde\Theta+\theta R}\right)}.
\end{align}
\end{Corollary}

\begin{IEEEproof}
The proof is given in Appendix \ref{sec:ProofSISORR}.
\end{IEEEproof}
Alternatively, we may (from the proof in Appendix \ref{sec:ProofSISORR}) solve $\mathrm{e}^{\theta R}={\lambda}^{-1}\mathrm{e}^{-\tilde\Theta(1-{\lambda}^{-1})}$ parametrically for the SNR $\Gamma$. By using (\ref{eq:Cefflambda}), we get the (more appealing) closed-form expression 
\begin{align}
\Gamma(C_\textrm{eff}^\textrm{RR})=\frac{(2^R-1)(\mathrm{e}^{\theta C_\textrm{eff}^\textrm{RR}}-1)}{\theta(R-C_\textrm{eff}^\textrm{RR})}.
\label{eq:GammaCeffPar}
\end{align}
Using (\ref{eq:GammaCeffPar}), we fix $R$ and can parametrically plot $C_\textrm{eff}^\textrm{RR}$ vs. $\Gamma$ as $(10\log_{10}(\Gamma(C_\textrm{eff}^\textrm{RR})),C_\textrm{eff}^\textrm{RR})$, for $0 \leq C_\textrm{eff}^\textrm{RR} \leq R$.

It is however more interesting, and easier, to consider the effective capacity for RR in a Rayleigh fading channel with respect to the QoS-parameter $\psi$. This is done in Corollary \ref{Corollary:CpsiRRRayleigh}.
\begin{Corollary} 
\label{Corollary:CpsiRRRayleigh} 
The effective capacity of persistent-RR in block Rayleigh fading channel, with $\tilde\Theta=(2^R-1)/\Gamma$, and $\psi\triangleq \log(\eta/\epsilon)/D_\textrm{max}$, is 
\begin{align}
C_\textrm{eff}^\textrm{RR}(\psi)
=\frac{R}{1+\tilde\Theta(\frac{\mathrm{e}^\psi-1}{\psi})}.
\label{eq:CpsiRRRayleigh}
\end{align}
\end{Corollary}
\begin{IEEEproof}
Using Corollary \ref{Corollary:CpsiMEDRR}, together with $F(s)=1/(1+s)$, allow us to write $a(s)=\mathrm{e}^\psi$ and  $b(s)=s+(1-\mathrm{e}^\psi)$. In the MED vector-matrix-form, this corresponds to $\mathbf{a}=\mathrm{e}^\psi$, $\mathbf{B}=1-\mathrm{e}^\psi$, $\mathbf{c}=1$, which leads to (\ref{eq:CpsiRRRayleigh}).
\end{IEEEproof}
We observe that (\ref{eq:CpsiRRRayleigh}), which takes the delay constraint $D_\textrm{max}$ and the delay violation probability $\epsilon$ into account via $\psi$, only involves an extra factor $(\mathrm{e}^\psi-1)/\psi$ compared to the  throughput for RR in Rayleigh fading, $T^\textrm{RR}=R/(1+\Theta)$, \cite[(12)]{bibLarsson1}. Since $\tilde \Theta\triangleq (2^R-1)/\Gamma$,  (\ref{eq:CpsiRRRayleigh}) is the same as the throughput \cite[(12)]{bibLarsson1}, but operating with a scaled SNR.

It is interesting to be able to compare (\ref{eq:CpsiRRRayleigh}) with the effective capacity of ARQ expressed in $\psi$. The following Corollary gives the effective capacity of ARQ.
\begin{Corollary}
\label{Corollary:CpsiARQ1}
The effective capacity of ARQ in block Rayleigh fading channel, with $\tilde\Theta=(2^R-1)/\Gamma$, and $\psi\triangleq \log(\eta/\epsilon)/D_\textrm{max}$, is 
\begin{align}
C_\textrm{eff}^\textrm{ARQ}(\psi)=\frac{R}{1+\psi^{-1}\ln \left(\frac{1-Q_1}{1-Q_1\mathrm{e}^\psi}\right)}, \ Q_1=1-\mathrm{e}^{-\tilde \Theta}
\label{eq:ARQCeffRpsi}
\end{align}
\end{Corollary}
\begin{IEEEproof}
Solve (\ref{CeffARQ}) for $\theta$ and use (\ref{eq:thetaC}). (Or use Corollary \ref{Corollary:CpsiHARQ1} with $P_m=P_1Q_1^{m-1}$). For Rayleigh fading, we also have $Q_1=1-\mathrm{e}^{-\tilde \Theta}$.
\end{IEEEproof}
At closer scrutiny, the RHS of (\ref{eq:ARQCeffRpsi}) converges to $R(1-Q_1)$, the throughput of ARQ, when $\psi \rightarrow 0$. Moreover, with a MED fading channel, $f_Z(z)=\mathbf{p}\mathrm{e}^{\Theta \mathbf{Q}}\mathbf{r}$, instead of Rayleigh fading, $Q_1$ in (\ref{eq:ARQCeffRpsi}) can be directly written as $Q_1=\mathbf{a}\mathrm{e}^{\Theta \mathbf{B}}\mathbf{c}$, where $\mathbf{a}=[0 \ \mathbf{p}]$, $\mathbf{b}=[0 \ \mathbf{q}]$, and $\mathbf{c}=[0; \mathbf{r}]$.

\begin{figure}[t]
 \centering
 \includegraphics[width=9cm]{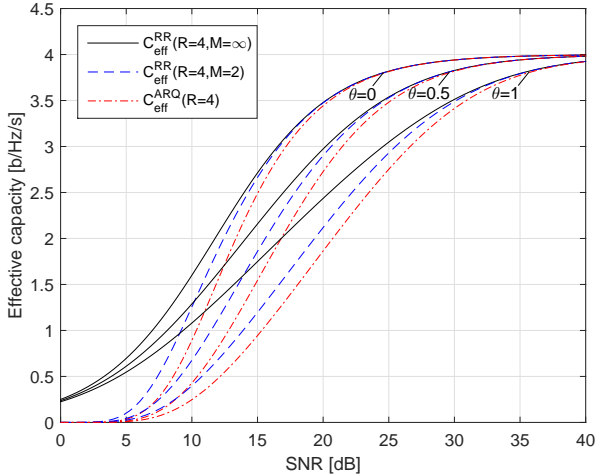}
 \vspace{-0.75cm}
 \caption{The effective capacity vs. SNR for ARQ, truncated-RR with $M=2$, and persistent-RR (with $M=\infty$), for $\theta=\{0,0.5,1\}$, and $R=4$.}
 \label{fig:CeffRR_M1_M2_Minf}
 \vspace{-0.25cm}
\end{figure}
\begin{figure}[t]
 \centering
 \includegraphics[width=9cm]{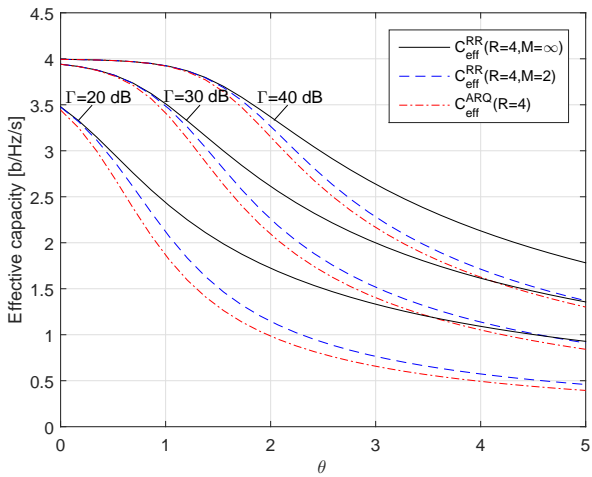}
 \vspace{-0.75cm}
 \caption{The effective capacity vs. $\theta$ for ARQ, truncated-RR with $M=2$, and persistent-RR (with $M=\infty$), for $\Gamma=\{20,30,40\}$ dB, and $R=4$.}
 \label{fig:CeffRRvstheta_M1_M2_Minf}
 \vspace{-0.25cm}
\end{figure}

\begin{figure}[!t]
 \centering
 \includegraphics[width=9cm]{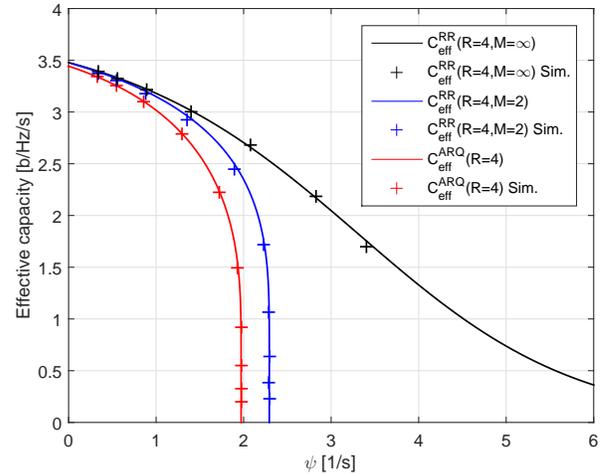}
 \vspace{-0.75cm}
 \caption{The effective capacity vs. $\psi$ for ARQ, truncated-RR with $M=2$, and persistent-RR (with $M=\infty$), for $\Gamma=\{20\}$ dB, and $R=4$. Analytical and Monte-Carlo simulated results are shown}
 \label{fig:CeffRR1}
 \vspace{-0.25cm}
\end{figure}

We are now prepared to compare the effective capacity of persistent-RR, truncated-RR with $M=2$, and ARQ. For simplicity, we consider a Rayleigh fading channel. In Fig. \ref{fig:CeffRR_M1_M2_Minf}, we plot the effective capacities of (\ref{eq:SISORRtheta}), (\ref{eq:Ceff2}),  and (\ref{CeffARQ}), for $\theta=\{0,0.5,1\}$ and $R=4$ b/Hz/s. It is noted that when the effective capacity is low, persistent-RR is more robust wrt $\theta$ than truncated-HARQ with $M=2$, as well as ARQ. However, when effective capacity is close to rate $R$, the schemes are similar in robustness to changes in $\theta$. This behavior is, at low SNR, due to maximum ratio combining over many transmissions for persistent-RR (and not the other two), and that all three schemes uses close to just one transmission when $C_\textrm{eff}\approx R$.
The same schemes are shown in Fig.  \ref{fig:CeffRRvstheta_M1_M2_Minf}, where the effective capacities are plotted vs. $\theta$ for $\Gamma=\{20,30,40\}$ dB and $R=4$ b/Hz/s. We note, as expected, that the effective capacity decreases with $\theta$. persistent-HARQ do not suffer the same performance loss when the QoS-requirement increases (with increasing $\theta$), since persistent-HARQ can benefit more from maximum ratio combining of multiple transmissions compared to the $M=2$ case. 
In Fig. \ref{fig:CeffRR1}, the same cases are studied, but for the effective capacities of (\ref{eq:CpsiRRRayleigh}), (\ref{eq:Ceff2}) with (\ref{eq:thetaC}), and (\ref{eq:ARQCeffRpsi}) vs. $\psi$. More interestingly, (\ref{eq:CeffNk}) with (\ref{eq:thetaC}) is Monte-Carlo simulated and compared to the analytical results, wherein we observe a perfect match.

\subsection{Network-Coded ARQ}
\label{sec:NCARQ}
So far, we have not yet demonstrated an example where the number of packets communicated during a transmission exceeds one, nor have we illustrated the use of multiple communication modes. In the following, we consider a three-mode example, where a reward of rate $2R$ occurs. Namely, we study the case of NC-ARQ for two users, user A and B, as described in \cite{bibLarsson3}. The main result is given in Corollary \ref{Corollary:NCARQ}.

\begin{Corollary}
\label{Corollary:NCARQ}
The effective capacity for 2-user NC-ARQ, with identical decoding probabilities $P_1$, $Q_1\triangleq 1-P_1$, and a common transmit queue, is given by Theorem \ref{theorem:CeffRetrScheme} for finite $k$ (or Corollary \ref{Corollary:CeffRetrSchemekInf} for infinite $k$) with
\begin{align}
\mathbf{A}=
\begin{bmatrix}
Q_1^2+P_1\mathrm{e}^{-\theta R} & 0 & P_1^2\mathrm{e}^{-\theta 2R}  \\
Q_1P_1 & Q_1^2+P_1\mathrm{e}^{-\theta R} &2P_1Q_1\mathrm{e}^{-\theta R}  \\
0 & P_1Q_1 & Q_1^2 
\end{bmatrix}.
\label{eq:A2NCARQ}
\end{align}
\end{Corollary}

\begin{IEEEproof}
For the analysis, it is (due to identical $P_1$, and hence symmetry) sufficient to consider a three-mode operation, with $s=\{1,2,3\}$. We start in communication mode $s=1$, which represents a normal ARQ operation. We enter mode $s=2$ when a data packet intended for user A is correctly decoded by user B, but not by user A. This occurs with probability $Q_1P_1$. In mode $s=2$, we transmit packets (in a normal ARQ fashion) for user B until user A, but not user B, decodes the packet correctly. At such event, we then enter mode $s=3$. This occurs with probability $P_1Q_1$. In mode $s=3$, we send the network-coded packet until it is either decoded by one of the users, and we then enter mode $s=2$ again with probability $2P_1Q_1$, or it is decoded by both users, and we then enter mode $s=1$ with probability $P_1^2$. For a more detailed description of NC-ARQ, e.g. with more users, we refer to \cite{bibLarsson3}. Following this described operation, the evolution of the communication modes is described by the following system of recurrence relations
\begin{align}
f_{k,1}&=(Q_1^2+P_1\mathrm{e}^{-\theta R})f_{k-1,1}+P_1^2\mathrm{e}^{-2\theta R}f_{k-1,3}, \label{eq:RecNCARQ1}\\
f_{k,2}&=P_1Q_1f_{k-1,1}+(Q_1^2+P_1\mathrm{e}^{-\theta R})f_{k-1,2}\notag\\
&+2P_1Q_1\mathrm{e}^{-\theta R}f_{k-1,3},\label{eq:RecNCARQ2}\\
f_{k,3}&=P_1Q_1 f_{k-1,2}+Q_1^2 f_{k-1,3}.
\label{eq:RecNCARQ3}
\end{align}
By ordering (\ref{eq:RecNCARQ1})-(\ref{eq:RecNCARQ3}) as a matrix recurrence relation, the matrix $\mathbf{A}$ in Corollary \ref{Corollary:NCARQ} is readily identified.
\end{IEEEproof} 
\begin{figure}[t]
 \centering
 \includegraphics[width=9cm]{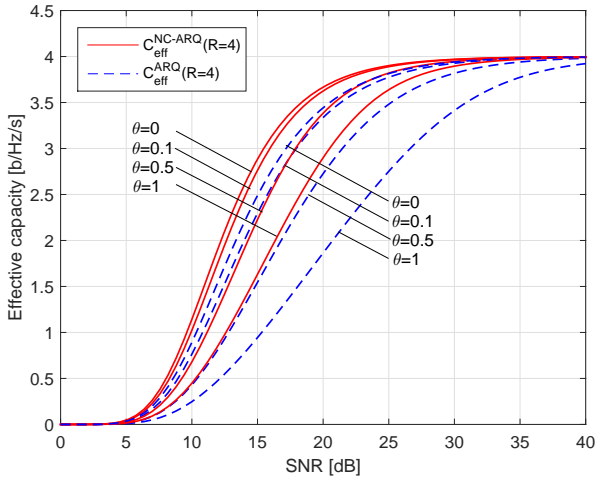}
 \vspace{-0.75cm}
 \caption{The effective capacity for NC-ARQ and ARQ with $R=4$ b/Hz/s for $\theta=\{0,0.1,0.5,1\}$.}
 \label{fig:CeffNCARQ}
 \vspace{-0.25cm}
\end{figure}
We now plot the effective capacity for the NC-ARQ matrix (\ref{eq:A2NCARQ}) and ARQ expression (\ref{eq:ARQCeffRpsi}) vs. SNR for different values of $\theta$ in Fig. \ref{fig:CeffNCARQ}. We note that NC-ARQ is less sensitive to an increase in $\theta$ (tougher QoS requirements), and provides higher effective capacity, than ARQ. This can be explained by the increased degree of diversity provided by NC-ARQ over ARQ. We expect this effect to increase with increasing number of users for NC-ARQ. The same phenomenon has been experimentally verified for effective channels with $F(s)=1/(1+s)^{\tilde N}$, where $\tilde N$ is the degree of diversity.
An obvious idea, based on the above, is to extend the analysis to more advanced joint network coding - retransmission schemes. However, it has proven hard (since \cite{bibLarsson3}) to design a structured, simple, and capacity achieving, scheme for NC-ARQ with more than 2-users. Nevertheless, with such design, the effective capacity can be determined with the framework proposed in this paper. Since, the framework handles all aspects of multiple transmissions, multiple communications modes, and multiple packet transmissions, we note that the framework is directly applicable to RR- and IR- based NC-HARQ described in \cite{bibLarsson4} if the transition probabilities, $P_{m\nu\tilde{s}s}$ are known. In Appendix \ref{sec:MultilayerARQ}, we present yet another retransmission scheme, based on superposition-coding and ARQ (extendible to HARQ), that can transfer multiple-packets concurrently.

\subsection{Two-mode ARQ and the Block Gilbert-Elliot Channel}
\label{sec:2modeARQ}
In this section, we illustrate that the developed framework also allows for analysis the combination of a retransmission scheme with a channel model that can change over time. Specifically, we consider two-mode ARQ, and operating in a block Gilbert-Elliot channel. As a first initial step, we analyze a more general system that alters between two modes with transition probabilities $P_{1\nu \tilde{s} s}, \tilde{s}=\{1,2\}, s=\{1,2\}$. The effective capacity is then given by the following Corollary.
\begin{Corollary}
\label{Corollary:TwoModeARQ}
The effective capacity for two-mode-ARQ, with transition probabilities $P_{1\nu \tilde{s} s}, \tilde{s}=\{1,2\}, s=\{1,2\}$, is computed with Corollary \ref{Corollary:CeffRetrSchemekInf} for infinite $k$ and 
\begin{align}
\lambda_+&=\frac{ (a_{111}+a_{122})+\sqrt{(a_{111}-a_{122})^2+4a_{121}a_{112}}}{2}
\label{eq:Ceff2mode},\\
&\text{where}\notag\\
a_{1 \tilde{s} s} &=\sum_{\nu=0}^{\nu_\emph{max}}P_{1\nu \tilde{s} s}\mathrm{e}^{-\theta \nu R}.
\label{eq:a1ss}
\end{align}
\end{Corollary} 

\begin{IEEEproof}
The two-mode (or two-state) matrix $\mathbf{A}$ is 
\begin{align}
\mathbf{A}=
\begin{bmatrix}
a_{111} & a_{121}\\
a_{112} & a_{122}
\end{bmatrix}.
\end{align}
We solve the characteristic equation of $\mathbf{A}$, $\lambda^2-\lambda(a_{111} +a_{122})+(a_{111}a_{122}-a_{121} a_{112})=0$, for $\lambda$.
\end{IEEEproof}
Specifically note the general form of $a_{1\tilde{s}s}$, \label{eq:a1ss}, i.e. a sum of several transition probabilities times a rate-dependent function, and again observe that this is more general than the entries in the $\mathbf{P}\boldsymbol{\Phi}$-form discussed in related works.

As discussed in Section \ref{sec:RetrSchemes}, the Gilbert-Elliot channel has been a popular tool to study ARQ system performance in channels with burst errors on symbol level. Here, we model a correlated block fading channel with packet (instead of symbol) errors. This could model a communication link that randomly (and with some correlation) shifts, e.g., communication media/frequency and hence propagation conditions. The probability of changing channel modes are denoted $\pi_{gg},\pi_{gb}=1-\pi_{gg},\pi_{bb},\pi_{bg}=1-\pi_{bb}$ for good-to-good, good-to-bad, bad-to-bad, and bad-to-good, respectively. We further assume that when operating in the good channel mode, the successful decoding probability is $p_g$, and the decoding failure probability is $q_g=1-p_g$. Similarly, in the bad channel mode, we have $p_b$, and  $q_b=1-p_b$, respectively. The effective capacity for the block Gilbert-Elliot channel is, due to independence between the transmission process and communication mode process, then given by $a_{111}=\pi_{gg}(q_g+p_g\mathrm{e}^{-\theta R})$, $a_{121}=\pi_{gb}(q_g+p_g\mathrm{e}^{-\theta R})$, $a_{122}=\pi_{bb}(q_b+p_b\mathrm{e}^{-\theta R})$, and $a_{112}=\pi_{bg}(q_b+p_b\mathrm{e}^{-\theta R})$ inserted in (\ref{eq:Ceff2mode}).

\section{Summary and Conclusions}
\label{sec:Conclusions}
In this paper, we have studied the effective capacity of general retransmission schemes, with multi-transmissions, -communication modes, and -packets. We modeled such schemes as a (constrained) random walk with transition probabilities, and the effective capacity could be compactly formulated and efficiently analyzed through a system of recurrence relations. From this, a matrix and a characteristic equation approach were developed to solve the recurrence(s). The characteristic equation method (with its simple form) turned out  to be particularly useful to gain new insights in many important cases, e.g. for truncated- and persistent-HARQ. This led to that many useful results could be formulated, in terms of general transition probabilities, general effective channel functions, or in specific wireless fading channels. With results expressed in a real QoS-metric dependent parameter, $\psi$, and the MED-channel, we could enhance the practical relevance even further. An interesting finding is that diversity, e.g. due to MRC or NC, or for that matter channels with small variance to mean, such as would be the case for spatially multiplexed MIMO, reduces the sensitivity of the effective capacity to $\theta$ (or to $\psi$). 
\appendix

\subsection{Asymptotic Convergence of the Effective Capacity}
\label{sec:CeffConverT}
The effective capacity (for $k$ timeslots) converges to the throughput (for $k$ timeslots) since
\begin{align}
&\lim_{\theta\rightarrow 0}C_{\textrm{eff},k}^\textrm{Retr.}
=-\lim_{\theta\rightarrow 0}\frac{1}{\theta k} \ln{\left(\mathbf{E}\{\mathrm{e}^{-\theta R N_k}\}\right)}\notag\\
&\overset{(a)}{=}\lim_{\theta\rightarrow 0}\frac{ 1}{ k}\left ( R \sum_{s=1}^S\sum_{\forall n} n\mathbb{P}\{N_k=n,S_k=s\}+\mathcal{O}(\theta)\right)\notag\\
&=\frac{R\mathbb{E}\{N_k\}}{ k}\triangleq T_k^\textrm{Retr.},
\end{align}
where we in step (a) used
\begin{align}
&\mathbf{E}\{\mathrm{e}^{-\theta R N_k}\}
=\sum_{s=1}^S\sum_{\forall n} \mathrm{e}^{-\theta R n}\mathbb{P}\{N_k=n,S_k=s\}\notag\\
&=\sum_{s=1}^S\sum_{\forall n} (1-\theta R n+\mathcal{O}(\theta^2))\mathbb{P}\{N_k=n,S_k=s\}\notag\\
&=1-\theta R \sum_{s=1}^S\sum_{\forall n} n\mathbb{P}\{N_k=n,S_k=s\}+\mathcal{O}(\theta^2)\notag ,
\end{align}
and exploited that $-\ln(1-x)\simeq x$ for small $x$.

\subsection{Throughput and $\alpha$-Moment of Finite-$k$ Truncated-HARQ}
\label{sec:kTSThroughput}
In this section, we consider a truncated-HARQ system with $k$ timeslots, and analyze the throughput in the Corollary below. Specifically, in the proof, we show that the recurrence relation idea is also applicable to compute the $\alpha$-moment $\mathbb{E}\{N_k^\alpha\}$.
\begin{Corollary}
\label{Corollary:THARQkSlot} 
The $k$-timeslot limited throughput, $T_k\triangleq R\mathbb{E}\{N_k\}/k$, of truncated-HARQ has the form
\begin{align}
T_{\textrm{trunc.},k}^\textrm{HARQ}&=T_\textrm{trunc.}^\textrm{HARQ}+\frac{1}{k}\sum_{m=1}^M c_m \lambda_m^k,
\label{eq:THARQrec}
\end{align}
where $\lambda_m$ are the solutions to the characteristic equation $\lambda^M=Q_M+\sum_{m=1}^MP_m\lambda^{M-m}$, and the constants $c_m$ are determined from initial conditions.
\end{Corollary}
\begin{IEEEproof}
First, a recurrence relation for the $\alpha$-moment can be expanded as
\begin{align}
&\mathbb{E}\{N_k^\alpha\}
=\sum_{\forall n} n^\alpha \mathbb{P}\{N_k=n\}=\sum_{\forall n} n^\alpha Q_M\mathbb{P}\{N_{k-M}=n\}\notag\\
&+\sum_{\forall n} ((n-1)+1)^\alpha \sum_{m=1}^M P_m\mathbb{P}\{N_{k-m}=n-1\}\notag\\
&= Q_M\mathbb{E}\{N_{k-M}^\alpha\}\notag\\
&+\sum_{\forall n} \sum_{m=1}^M P_m \sum_{\beta=0}^\alpha { \alpha \choose \beta}(n-1)^\beta\mathbb{P}\{N_{k-m}=n-1\}\notag\\
&= Q_M\mathbb{E}\{N_{k-M}^\alpha\}\notag\\
&+\sum_{m=1}^M  \sum_{\beta=1}^\alpha P_m { \alpha \choose \beta}\mathbb{E}\{N_{k-m}^{\beta}\}
+1-Q_M.
\label{eq:AlphaMoment}
\end{align}
Now, consider (\ref{eq:AlphaMoment}) with $\alpha=1$, multiply both sides with $R$, extend the $(k-m)$th mean-terms with $(k-m)/(k-m)$, and use the definition $T_k\triangleq R\mathbb{E}\{N_k\}/k$. We then arrive at the recurrence relation $kT_k=Q_M(k-M)T_{k-M}+\sum_{m=1}^M P_m(k-m)T_{k-m}+R(1-Q_M)$. Now insert $T_k=T^\textrm{HARQ}_\textrm{trunc.}+t_k$, with $T^\textrm{HARQ}_\textrm{trunc.}$ from (\ref{eq:THARQtrunc}), in the recurrence relation which is then transformed into the homogeneous form
$kt_k=Q_M(k-M)t_{k-M}+\sum_{m=1}^M P_m(k-m)t_{k-m}$.
Then with $h_k=kt_k$, we get $h_k=Q_M h_{k-M}+\sum_{m=1}^M P_m h_{k-m}$, which has the general solution $h_k=\sum_{m=1}^M c_m \lambda_m^k$ if all roots are unique.
\end{IEEEproof}
Note that higher moments from (\ref{eq:AlphaMoment}) can be useful elsewhere. For example, the approximate effective capacity, for $k$ timeslots and small $\theta$, is $C_{\textrm{eff},k}^\textrm{HARQ}=-\ln(1-\theta R \mathbb{E}\{N_k\}+\theta^2 R^2 \mathbb{E}\{N_k^2\}/2+\mathcal{O}(\theta^3))/\theta k\approx
T_{\textrm{trunc.},k}^\textrm{HARQ}-\theta R^2 \mathbb{E}\{N_k^2\}/2k$.

\subsection{Effective Capacity Analysis of Truncated-HARQ}
\label{ECHARQApp}
For the readers convenience, we give the (more tractable) analysis of the effective capacity for truncated-HARQ. The corresponding recurrence relation is given by
\begin{align}
&\mathbb{E}\left\{ \mathrm{e}^{-\theta R N_k}\right \} 
= \sum_{\forall n}\mathrm{e}^{-\theta R n} \mathbb{P}\left\{ N_k=n\right \} \notag\\
&\overset{(a)}{=} \sum_{\forall n}\mathrm{e}^{-\theta R n} \left( \sum_{m=1}^M P_{m}\mathbb{P}\left\{ N_{k-m}=n-1\right \} \right. \notag\\
&+ Q_{M}\mathbb{P}\left\{ N_{k-M}=n\right \}  \Bigg)\notag\\
&= \sum_{m=1}^M  P_{m}\mathrm{e}^{-\theta R}  \sum_{\forall n}\mathrm{e}^{-\theta R (n-1)}\mathbb{P}\left\{ N_{k-m}=n-1\right \} \notag\\
& +Q_{M}\sum_{\forall n}\mathrm{e}^{-\theta R n}\mathbb{P}\left\{ N_{k-M}=n\right \} \notag\\
&= \sum_{m=1}^M   P_{m}\mathrm{e}^{-\theta R} \mathbb{E}\left\{ \mathrm{e}^{-\theta R N_{k-m}}\right \}+Q_M \mathbb{E}\left\{ \mathrm{e}^{-\theta R N_{k-M}}\right \},
\label{eq:ENkHARQ}
\end{align}
where the identity (\ref{eq:P}) is used in step (a). For the identity, we exploit that $\pi_{k,n}=\sum_{m=1}^M P_{m} \pi_{k-m,n-1}+Q_M\pi_{k-M,n}$, and we let $d_0=1$, $d_\mu=\sum_{j=m+1}^{M} P_j, \mu\in \{1,2,\ldots, M-1\}$, which yields
\begin{align}
&\mathbb{P}\left\{ N_{k}=n\right \} 
= \sum_{\mu=0}^{M-1} d_\mu\pi_{k-\mu,n}\notag\\
&= \sum_{\mu=0}^{M-1} d_\mu\left (\sum_{m=1}^M P_m \pi_{k-\mu-m,n-1}  +Q_M\pi_{k-\mu-M,n} \right )\notag\\
&=\sum_{m=1}^M P_m   \sum_{\mu=0}^{M-1} d_\mu\pi_{(k-m)-\mu,n-1}
\notag\\
&+Q_M   \sum_{\mu=0}^{M-1} d_\mu\pi_{(k-M)-\mu,n}\notag\\
&=\sum_{m=1}^M P_m  \mathbb{P}\left\{ N_{k-m}=n-1\right \}+Q_M  \mathbb{P}\left\{ N_{k-M}=n\right \}.
\label{eq:P}
\end{align}

\subsection{Proof of Corollary \ref{Corollary:Valid1}}
\label{sec:ProofValid1}
\begin{IEEEproof} Descarte's rule of signs in algebra states that the largest number of positive roots to a polynomial equation with real coefficients and ordered by descending variable exponent is upper limited by the number of sign change. Since all $a_m>0$, we only have one sign change, and thus just one positive root. In algebra, it is known that if a function $f(x)$ has at least one positive root in the interval $[a,b]$, then $f(a)f(b)<0$. Let $f(\lambda)$ be the characteristic polynomial, we then have $f(0)f(1)=(-a_M)(1-\mathrm{e}^{-\theta R})(\sum_{m=1}^M P_m)<0$. Thus, we have one positive root, and it lies in the interval $[0,1]$.\end{IEEEproof}

\subsection{Proof of Corollary \ref{Corollary:ECtoTHARQ}}
\label{sec:ProofValid2}
\begin{IEEEproof}
We rewrite  (\ref{eq:Cefflambda}) as $\lambda_+=\mathrm{e}^{-\theta C_\textrm{eff}^\textrm{HARQ}}$, which is inserted into the characteristic equation (\ref{eq:HARQCharEq}), that then gives
\begin{align}
\mathrm{e}^{-M \theta C_\textrm{eff}^\textrm{HARQ}}&=\mathrm{e}^{-\theta R}\sum_{m=1}^M P_m\mathrm{e}^{-(M-m)\theta C_\textrm{eff}^\textrm{HARQ}}+Q_M\notag\\
\Rightarrow \mathrm{e}^{\theta R}&=\sum_{m=1}^M P_m\mathrm{e}^{m\theta C_\textrm{eff}^\textrm{HARQ}}+Q_M\mathrm{e}^{\theta (R+M C_\textrm{eff}^\textrm{HARQ})}\notag\\
\overset{(a)}{\Rightarrow}  (1+\theta R)&\approx\sum_{m=1}^M P_m(1+m\theta C_\textrm{eff}^\textrm{HARQ})\notag\\
&+Q_M(1+\theta R+M\theta C_\textrm{eff}^\textrm{HARQ})\notag\\
\overset{(b)}{\Rightarrow}  R(1-Q_M)&=\sum_{m=1}^M m P_m  C_\textrm{eff}^\textrm{HARQ}+Q_M M C_\textrm{eff}^\textrm{HARQ} \notag\\
\Rightarrow  \lim_{\theta\rightarrow 0}C_\textrm{eff}^\textrm{HARQ} &=\frac{R(1-Q_M)}{\sum_{m=1}^M m P_m+MQ_M }.
\label{eq:CeffTHARQproof}
\end{align}
For (\ref{eq:CeffTHARQproof}), in step (a) we expanded for small $\theta$, and in step (b), we used the fact that $\sum_{m=1}^M P_m+Q_M=1$.
\end{IEEEproof}

\subsection{Proof of Corollary \ref{Corollary:ECGeomARQ}}
\label{sec:ProofValid3}
\begin{IEEEproof}
The characteristic equation, with the truncated geometric probability mass function inserted, is solved for $\lambda_+$ below. We find that
\begin{align}
\lambda^M&=\sum_{m=1}^M P_1Q_1^{m-1}\mathrm{e}^{-\theta R}\lambda^{M-m}+Q_1^M\notag\\
\Rightarrow \left(\frac{\lambda}{Q_1}\right)^M-1&=
\frac{P_1\mathrm{e}^{-\theta R}}{Q_1}\sum_{m=1}^M \left(\frac{\lambda}{Q_1}\right)^{M-m}\notag\\
\Rightarrow \left({\lambda}/{Q_1}\right)^M-1&=
\frac{P_1\mathrm{e}^{-\theta R}}{Q_1}\frac{(\lambda/Q_1)^M-1}{(\lambda/Q_1)-1}\notag\\
\overset{(a)}{\Rightarrow} \lambda_+ &= P_1\mathrm{e}^{-\theta R}+Q_1,
\end{align}
where the step (a) is due to that only one root exists.
\end{IEEEproof}

\subsection{Proof of Corollary \ref{Corollary:CeffAppr}}
\label{sec:ProofAppl2}
\begin{IEEEproof}
Consider (\ref{eq:HARQCharEq}), divide by $\lambda^M$, insert $\lambda=\mathrm{e}^{-\theta C_\textrm{eff}^\textrm{HARQ}}$, and let $M\rightarrow \infty$. We then get
\begin{align}
 1&=\sum_{m=1}^\infty P_m\mathrm{e}^{\theta( m C_\textrm{eff}^\textrm{HARQ}-R)}\notag\\
&\overset{(a)}{\approx}\sum_{m=1}^\infty P_m(1+\theta (mC_\textrm{eff}^\textrm{HARQ}-R)+\theta^2(mC_\textrm{eff}^\textrm{HARQ}-R)^2/2)\notag\\
&\overset{(b)}{\Rightarrow}   (C_\textrm{eff}^\textrm{HARQ})^2+ c_1C_\textrm{eff}^\textrm{HARQ}-\tilde{c}_2=0,
\end{align}
where we Taylor-expanded for small $\theta$ in the step (a), and rewrote the expression as a quadratic equation in the step (b) with $c_1, c_2$ being defined as in Corollary \ref{Corollary:CeffAppr}. We then solve the quadratic equation for $C_\textrm{eff}^\textrm{HARQ}$.
\end{IEEEproof} 

\subsection{Proof of Corollary \ref{Corollary:SISORR}}
\label{sec:ProofSISORR}
Here, we prove the effective capacity expression for RR-HARQ in Rayleigh fading expressed in  $\theta$.
\begin{IEEEproof}
Using Theorem \ref{theorem:ECHARQEffCh} with $F(s)=1/(1+s)$, we get the expression
\begin{align}
\mathrm{e}^{\theta R}&= \mathcal{L}_{\tilde\Theta} ^{-1} \left \{ \frac{1}{s}\frac{1-\frac{1}{1+s}}{\lambda-\frac{1}{1+s}} \right \}\notag\\
\Rightarrow \mathrm{e}^{\theta R}&=\frac{1}{\lambda}\mathrm{e}^{-\Theta(1-\frac{1}{\lambda})}\notag\\
\Rightarrow \lambda_+&=\frac{\tilde\Theta}{W_0 \left( \tilde\Theta \mathrm{e}^{\tilde\Theta+\theta R}\right)}.
\end{align}
Taking the logarithm of $\lambda_+$, a more compact form is
\begin{align}
 \ln(\lambda_+)&=\ln \left( \frac{\tilde\Theta}{W_0 \left( \tilde\Theta \mathrm{e}^{\tilde\Theta+\theta R}\right)}\right)\notag\\
&\overset{(a)}{=}\ln \left( \frac{\mathrm{e}^{W_0 \left( \tilde\Theta \mathrm{e}^{\tilde\Theta+\theta R}\right)}}{ \mathrm{e}^{\tilde\Theta+\theta R}}\right)\notag\\
&=W_0 \left( \tilde\Theta \mathrm{e}^{\tilde\Theta+\theta R}\right)-\tilde\Theta-\theta R,
\end{align}
which is inserted in (\ref{eq:Cefflambda}). For step (a), we used the definition of Lambert's-$W$ function, $x/W_0(x)=\mathrm{e}^{W_0(x)}$.
\end{IEEEproof}

\subsection{Optimized Effective Capacity of ARQ in Rayleigh Fading}
\label{sec:SISO-ARQ}
In several past works, e.g. \cite{bibBettesh}, \cite{bibSzczecinski}-\cite{bibLarsson2}, the rate-optimized throughput was examined for block Rayleigh fading, having $P_1=\mathrm{e}^{-\Theta}$, $\Theta=(2^R-1)/\Gamma$. In contrast, although the exact effective capacity expression is known for ARQ (\ref{CeffARQ}), the rate-optimized effective capacity expression is not. The (likely) reason for this is that the problem is (seemingly) intractable. Adopting the parametric optimization method explored in \cite{bibLarsson1} for throughput optimization, we now illustrate its use for effective capacity optimization below.
\begin{Corollary}
\label{Corollary:CeffOptARQ}
The maximum effective capacity of ARQ in block Rayleigh fading, vs. the optimal rate $R^*$, is
\begin{align}
\Gamma(R^*)&=\frac{\ln(2)2^{R^*}(\mathrm{e}^{ \theta R^*}-1)}{ \theta},\label{eq:ARQgammaR}\\
P_1(R^*)&=\mathrm{e}^{-\Theta}, \, \Theta(R^*)=\frac{2^{R^*}-1}{\Gamma},  \label{eq:ARQProbR}\\
C_\textrm{eff}^\textrm{ARQ*}(R^*)&=-\frac{1}{\theta} \ln \left(Q_1+P_1\mathrm{e}^{-\theta R^*}\right), \, Q_1\triangleq1-P_1.\label{eq:ARQCeffR}  
\end{align}
\end{Corollary}

\begin{figure}[t]
 \centering
 \includegraphics[width=9cm]{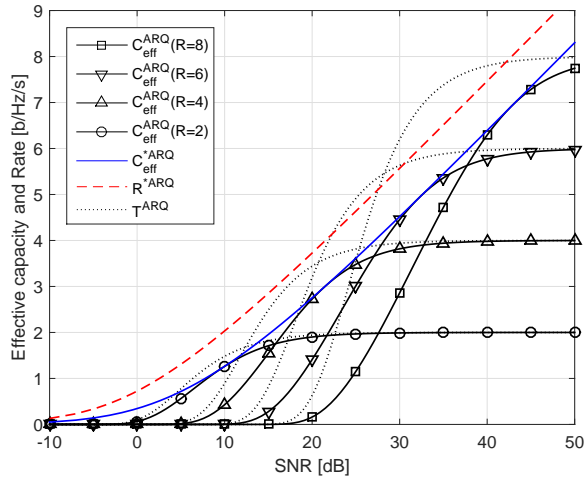}
 \vspace{-0.75cm}
 \caption{The effective capacity with $\theta=0.5$ (and throughput with $\theta=0$) for $R=\{2,4,6,8\}$, as well as the optimal effective capacity and optimal rate point are shown vs. SNR for ARQ.}
 \label{fig:CeffARQ2}
 \vspace{-0.25cm}
\end{figure}

\begin{IEEEproof}
We see in (\ref{CeffARQ}) that to maximize $C_\textrm{eff}^\textrm{ARQ}$, we can instead minimize $P_1(\mathrm{e}^{-\theta R}-1)$. Taking the derivative with respect to $R$, equating to zero, it is easy to show that the optimality criteria is $\Gamma  \theta \mathrm{2}^{-R}/\ln(2)-\mathrm{e}^{ \theta R}+1=0$, and additional checks ensures a minimum for positive $R$. While it is hard to solve for $R^*(\Gamma)$ into a closed-form expression, it is straightforward to solve for $\Gamma(R^*)$ instead. This is doable if a (monotonic) one-to-one mapping exists between $R^*$ and $\Gamma$. We then get $\Gamma(R^*)=\ln(2)\mathrm{e}^{R^*}(\mathrm{e}^{ \theta R^*}-1)/ \theta$, which is inserted in $\Theta=(2^R-1)/\Gamma$, which then gives $P_1=\mathrm{e}^{-\Theta}$ and $Q_1$. The latter two are subsequently inserted in $C_\textrm{eff}^\textrm{ARQ}$.
\end{IEEEproof}
The effective capacity expression of ARQ is plotted vs. SNR in Fig. \ref{fig:CeffARQ2} with $R=\{2,4,6,8\}$ b/Hz/s and $\theta=0.5$. We also show the parametrized maximum effective capacity value and the optimal rate point (\ref{eq:ARQgammaR})-(\ref{eq:ARQCeffR}) for ARQ and confirm their correctness. 

\subsection{Multilayer-ARQ}
\label{sec:MultilayerARQ}
With NC-ARQ in Section \ref{sec:NCARQ}, we presented a retransmission protocol which may concurrently communicate multiple packets in a single transmission. Another multi-packet retransmission approach is multilayer-ARQ \cite{bibLarsson5}. The idea is to concurrently transmit multiple super-positioned codewords, each with the SNR level $x_l \Gamma$, fractional power allocation $x_l$, mean SNR  $\Gamma$, rate $r_l$ for "level" $l\in\{1,2,\ldots,L\}$, over a block fading channel and decode as many codewords as possible. With proper power and rate allocation, it has been found that the throughput of such schemes exceeds traditional (single-layer) ARQ. The effective capacity of multilayer-ARQ is simply
\begin{align}
C_\textrm{eff}^\textrm{LARQ}&=- \frac{1}{\theta}\ln \left(q+\sum_{l=1}^L p_l\mathrm{e}^{-\theta R_l}\right)\notag\\
&=- \frac{1}{\theta}\ln \left(1+\sum_{l=1}^L p_l \left(\mathrm{e}^{-\theta R_l}-1\right)\right),
\label{eq:CeffMultiLayerARQ}
\end{align}
where $q=1-\sum_{l'=1}^L p_{l'}$ is the probability of decoding no packet, and $R_l=\sum_{l'=1}^l r_{l'}$ is the cumulative rate split. The probabilities of successful decoding for up to layer $l$ is
\begin{align}
p_l&=\mathbb{P}\left\{\log_2\left(1+\frac{(1-X_{l+1})\Gamma g}{1+X_{l+1}\Gamma g} \right)>R_l\right\}
=\mathbb{P}\left\{g >\Theta_l Y_l\right\}
\end{align}
where $X_l\triangleq \sum_{l'=l}^Lx_{l'}$ is the cumulative fractional power split, $Y_l\triangleq 1/(1-2^{R_l}X_{l+1})$ is a function of the power split, and $\Theta_l\triangleq ({2^{R_l}-1})/\Gamma$. For block Rayleigh fading, we have $p_l=\mathrm{e}^{-\Theta_l Y_l}$.  For multilayer-ARQ, exactly like for the throughput metric, analytical joint power-and-rate optimization of the effective capacity is intractable. Limiting the per layer rate to $r_l=r, \forall l$, multilayer-HARQ, and joint network coding and multilayer-ARQ, should be possible to handle within the analytical framework of this paper by taking advantage of multiple communication modes. We also note that while (\ref{eq:CeffMultiLayerARQ}) has a similar form as \cite[(25),(26)]{bibTang1}, but the latter does not consider multilayer-ARQ, but changes constellation sizes based on the current SNR in a block-fading channel.

\vspace{-0.0cm}
\bibliographystyle{IEEEbib}
\bibliography{strings,refs,manuals}


\end{document}